\documentclass[a4paper,11pt]{article} 
\usepackage{jheppub}
\usepackage{ifpdf}
\usepackage{graphicx,subcaption}
\usepackage{amsfonts}
\usepackage{amsmath}
\usepackage{amssymb}
\usepackage{pdfpages}
\usepackage{epstopdf}
\usepackage[makeroom]{cancel}
\usepackage{array}
\usepackage[normalem]{ulem}
\usepackage{xspace}

\newcommand{\beq}{\begin{equation}}
\newcommand{\eeq}{\end{equation}}

\newcommand{\reef}[1]{(\ref{#1})}

\newcommand{\mt}[1]{\textrm{\tiny #1}}

\newcommand{\cv}{{\cal C}_\mt{V}}
\newcommand{\cvcorr}{{\cal C}_\mt{corr}}

\newcommand{\UR}{U_\mt{R}}
\newcommand{\UL}{U_\mt{L}}
\newcommand{\tR}{t_\mt{R}}
\newcommand{\tL}{t_\mt{L}}

\begin{document}
\preprint{arXiv:2604.15730 [hep-th]}
\title{Stringy Effects on Holographic Complexity: The Complete Volume in Dynamical Spacetimes}
\author[a,b]{Qi Yang}
\author[a,b,1]{and Yu-Xiao Liu\note{Corresponding author.}}

\affiliation[a]{Lanzhou Center for Theoretical Physics, 
  Key Laboratory of Theoretical Physics of Gansu Province, 
  Key Laboratory of Quantum Theory and Applications of MoE, 
  Gansu Provincial Research Center for Basic Disciplines of Quantum Physics,
  Lanzhou University, Lanzhou 730000, China}
\affiliation[b]{Institute of Theoretical Physics \& Research Center of Gravitation,
  School of Physical Science and Technology, 
  Lanzhou University, Lanzhou 730000, China}

\emailAdd{yqi2024@lzu.edu.cn}
\emailAdd{liuyx@lzu.edu.cn}

\abstract{We investigate the stringy effects on holographic complexity in $(d+1)$-di\-men\-sion\-al Gauss-Bonnet gravity using the ``complete volume'' proposal for higher-curvature theories. 
Our analysis covers unperturbed eternal black holes, as well as the one-sided and two-sided Vaidya spacetimes. The one-sided geometry describes a null shell collapsing into the empty AdS vacuum to form a black hole, while the two-sided geometry represents a null shell injected into an eternal black hole background with arbitrary energy. For unperturbed backgrounds, higher-curvature terms introduce explicit corrections to the standard CV proposal, giving rise to a ``competition effect'' absent in the uncorrected framework. In the dynamical settings, we demonstrate that despite novel jumps in the canonical velocities across the null shell, the complexity growth rate remains universally governed by the conserved momentum, just as in Einstein gravity. Furthermore, our two-sided shock wave analysis reveals that Gauss-Bonnet corrections prolong the critical time, preserving the universal logarithmic dependence for the scrambling time.}

\maketitle 
\section{Introduction}\label{sec:Intro}

In recent years, the growing intersection of quantum information and quantum gravity has yielded profound insights, particularly within the framework of the AdS/CFT correspondence~\cite{Maldacena:1997re}. A seminal realization of this connection is the Ryu-Takayanagi (RT) formula and its covariant generalizations~\cite{Ryu:2006bv,Ryu:2006ef,Hubeny:2007xt,Rangamani:2016dms}, which relate the boundary entanglement entropy to bulk geometry, thereby demonstrating how semi-classical spacetime emerges from quantum entanglement~\cite{VanRaamsdonk:2009ar,VanRaamsdonk:2010pw}. This correspondence extends well beyond standard asymptotically AdS backgrounds. In holographic setups with a finite radial cutoff, such as those dual to irrelevant $T\bar{T}$ deformations~\cite{Smirnov:2016lqw,Cavaglia:2016oda,He:2025ppz}, RT-like prescriptions continue to provide crucial consistency checks of the duality~\cite{McGough:2016lol,Kraus:2018xrn,Chang:2024voo,Chang:2025ays}.

However, holographic entanglement entropy is insufficient to capture the late-time dynamical evolution of the black hole interior~\cite{Susskind:2014moa}. This limitation motivated the introduction of quantum computational complexity~\cite{Aaronson:2016vto} into the holographic dictionary. To describe the quantum complexity of boundary states, two pioneering holographic conjectures established the foundation: the complexity=volume (CV)~\cite{Susskind:2014rva,Stanford:2014jda} and complexity=action (CA)~\cite{Brown:2015bva,Brown:2015lvg} proposals. These proposals have subsequently inspired further generalizations, including the spacetime volume (CV 2.0)~\cite{Couch:2016exn} and the complexity=anything (CAny)~\cite{Belin:2021bga,Jorstad:2023kmq,Belin:2022xmt,Myers:2024vve}.

The CV proposal states that the complexity of a boundary state defined on a time slice $\Sigma$ is dual to the volume of an extremal codimension-one bulk hypersurface $\mathcal{B}$ anchored at $\Sigma$,
\beq
\cv(\Sigma)=\max _{\Sigma=\partial \mathcal{B}}\left[\frac{V(\mathcal{B})}{G_{\mathrm{bulk}} \ell}\right] \,,
\label{CV}
\eeq
where $G_{\mathrm{bulk}}$ and $\ell$ denote the bulk Newton constant and some undetermined length scale, respectively. While the complexity=action (CA) proposal natively incorporates higher-curvature modifications via the bulk action, determining the requisite surface and joint terms for general higher-derivative theories poses highly non-trivial technical challenges~\cite{Lehner:2016vdi,Cano:2018aqi,Cano:2018ckq,Chakraborty:2018dvi,Jiang:2018sqj}. Given these ambiguities, we explicitly set aside action-related proposals in this work, directing our focus entirely toward volume-based complexity proposals.

To gain deeper insights into the dynamical properties of complexity, investigations have been extended to time-dependent geometries~\cite{Frey:2021cjs}, particularly to Vaidya spacetimes~\cite{Vaidya:1999zz,Vaidya:1951zz,Wang:1998qx}. Numerous studies have examined the complexity evolution for one-sided black holes formed by a null shell collapsing into the AdS vacuum~\cite{Chapman:2018dem,Moosa:2017yvt,Susskind:2014jwa}, as well as for two-sided black holes perturbed by shock waves falling into an existing eternal black hole~\cite{Roberts:2014isa,Stanford:2014jda,Chapman:2018lsv}. In these two-sided shock wave geometries, holographic calculations have successfully reproduced characteristic features of quantum circuits, such as the switchback effect and the scrambling time, which are essential for understanding quantum chaos and information spreading in holographic systems. Furthermore, they reveal distinct evolutionary phases, including a transient plateau where the growth rate is proportional to the mass difference~\cite{Chapman:2018lsv}.

However, the majority of these explorations have been restricted to Einstein gravity. In the framework of string theory, the low-energy effective action is naturally supplemented by an infinite tower of $\alpha'$ corrections originating from worldsheet loop diagrams~\cite{Zwiebach:1985uq,Zumino:1985dp,Gross:1986iv,Tong:2009np}. Specifically, the standard Einstein-Hilbert action is generalized to include higher-curvature terms, taking the form
\begin{equation}
    S = \frac{1}{16\pi G_{\mathrm{bulk}}} \int d^{d+1} x \sqrt{-g} \left[ R - 2\Lambda + \sum_{n=1}^\infty \alpha'^n \mathcal{L}_n \right],
\end{equation}
where $\Lambda$ is the cosmological constant, and $\mathcal{L}_n$ are $n$-th order polynomials constructed from the Riemann tensor and potentially gauge fields for non-neutral backgrounds. To tractably capture the leading stringy effects, it is customary to focus on the canonical higher-curvature theory: Gauss-Bonnet (GB) gravity, which supplements the Lagrangian with a specific curvature term: $\mathcal{L} \supset \alpha' (R_{\mu\nu\rho\sigma}R^{\mu\nu\rho\sigma} - 4R_{\mu\nu}R^{\mu\nu} + R^2)$. This precise combination not only emerges naturally from many string compactifications~\cite{Sen:2005iz} but also belongs to the Lovelock family~\cite{Lovelock:1971yv}. Investigating these stringy effects is physically compelling. Historically, while the GB term famously violates the universal Kovtun-Starinets-Son viscosity bound~\cite{Brigante:2007nu}, the allowable magnitude of such higher-curvature corrections is strictly constrained by the requirements of causality~\cite{Brigante:2008gz,Buchel:2009tt} and the positivity of energy condition~\cite{Hofman:2009ug} in the dual conformal field theory. It is therefore crucial to understand how these stringy corrections alter the properties of holographic complexity~\cite{Frey:2024tnn}.

While holographic complexity in GB gravity has been explored via standard CV, CAny, and other tailored proposals~\cite{An:2018dbz,Nally:2019rnw, Wang:2023noo,Jiang:2023jti,Emami:2024mes}, evaluating higher-curvature corrections requires modifications. This requirement is best illustrated through an analogy with black hole entropy. Just as the Bekenstein-Hawking entropy~\cite{Bekenstein:1973ur,Bardeen:1973gs,Hawking:1975vcx}
\begin{equation}
    S_{\mathrm{BH}}= \frac{\text{Area}}{4G_{\mathrm{bulk}}}
\end{equation}
must be replaced by the Wald entropy~\cite{Wald:1993nt,Iyer:1994ys,Jacobson:1993vj} in theories of higher-derivative gravity, a similar generalization is required for the Ryu-Takayanagi prescription. Unlike Killing horizons, minimal surfaces generally have nonzero extrinsic curvature. To account for this, the holographic entanglement entropy must incorporate Dong's corrections~\cite{Dong:2013qoa}. Naturally, the CV proposal requires analogous modifications~\cite{Alishahiha:2017hwg,An:2018dbz,Nally:2019rnw,Jiang:2019kks}.
To rigorously address this, a generalization of the CV proposal was recently derived~\cite{Hernandez:2020nem}, motivated by entanglement wedge reconstruction and the island rule~\cite{Chen:2020uac,Chen:2020hmv}. This ``complete volume'' proposal introduces a distinct feature: incorporating these Wald-like geometric corrections yields an effective Lagrangian that inherently contains second derivatives, necessitating advanced variational techniques~\cite{Belin:2022xmt}. Unlike in the CAny framework, where second-derivative terms may arise from an arbitrary choice of bulk observables, here they emerge natively from the geometric modifications to the volume functional itself. Originally formulated for subregion complexity on the brane, this corrected complete volume functional is naturally extended by us to evaluate the global complexity. To guarantee a well-posed variational principle, the geometric functional must be supplemented with a boundary term $W_{\mathrm{bdy}}$. Consequently, the complete volume functional is given by
\beq
\cvcorr(\tau)=\max _{\partial \mathcal{B}(\tau)=\mathcal{B}_\tau}\left[\frac{W_{\mathrm{gen}}(\mathcal{B})+W_K(\mathcal{B})+W_{\mathrm{bdy}}\left(\partial \mathcal{B} \right)}{G_{\mathrm{bulk}} \ell}\right] , 
\label{complete_volume_functional}
\eeq
In a similar spirit to the Wald-Dong entropy, the hypersurface portion of this functional contains two main contributions. The first is a generalized volume $W_{\mathrm{gen}}$, which smoothly reduces to the standard volume $V(\mathcal{B})$ in the limit of Einstein gravity. It was initially conjectured with undetermined coefficients~\cite{Bueno:2016gnv} that were subsequently fixed by carefully examining the higher-curvature corrections via the Fefferman-Graham expansion near a brane. The second contribution, $W_K$, introduces an additional correction involving the extrinsic curvature $\mathcal{K}_{\mu\nu}$ of the hypersurface $\mathcal{B}$, analogous to Dong's corrections to the Wald entropy~\cite{Dong:2013qoa}. These corrections naturally arise from matching the subleading terms in this near-brane expansion. To evaluate these contributions explicitly, we use $x^\mu$ to denote the bulk coordinates, and use $\sigma^\alpha = (z, \sigma^a)$ to denote Gaussian normal coordinates on $\mathcal{B}$. For a general bulk Lagrangian $\mathbf{L}_{\mathrm{bulk}}$, these two bulk contributions evaluate to~\cite{Hernandez:2020nem}
\begin{equation}
\begin{aligned}
W_\mathrm{gen}(\mathcal{B}) &= \frac{2}{(d-1)(d-2)}\int_{\mathcal{B}}\mathrm{d}^{d}\sigma\sqrt{\det h}\left(1+(d-3)\frac{\partial\mathbf{L}_{\mathrm{bulk}}}{\partial\mathcal{R}_{\mu\nu\rho\sigma}}n_{\mu}h_{\nu\rho}n_{\sigma}\right) \,, \\
W_K(\mathcal{B}) &= \frac{4(d-3)}{(d-1)^{2}(d-2)}\int_{\mathcal{B}}\mathrm{d}^{d}\sigma\sqrt{\det h}\frac{\partial^{2}\mathbf{L}_{\mathrm{bulk}}}{\partial\mathcal{R}_{\mu_{1}\nu_{1}\rho_{1}\sigma_{1}}\partial\mathcal{R}_{\mu_{2}\nu_{2}\rho_{2}\sigma_{2}}} \\
 & \quad \times \left[\mathcal{K}_{\nu_{1}\sigma_{1}}(h_{\mu_{1}\rho_{1}}+(d-2)n_{\mu_{1}}n_{\rho_{1}})\mathcal{K}_{\nu_{2}\sigma_{2}}(h_{\mu_{2}\rho_{2}}+(d-2)n_{\mu_{2}}n_{\rho_{2}})\right]+\cdots \,.
 \label{eq:W_definitions}
\end{aligned}
\end{equation}
Here, $n^\mu$ and $h_{\mu\nu}$ denote the unit normal vector and the induced metric of the codimension-one spacelike hypersurface $\mathcal{B}$, respectively. In these expressions, the gravitational Lagrangian is rescaled such that the gravitational action carries an overall factor $I_\mathrm{grav}=\frac{1}{16\pi G_{\mathrm{bulk}}}\int \mathrm{d}^{d+1}x\sqrt{-g}\,\mathbf{L}_{\mathrm{bulk}}.$ 

The structural parallelism between the complexity functional~\eqref{eq:W_definitions} and the Wald-Dong entropy is explicit. For direct comparison, Dong's covariant holographic entanglement entropy evaluates on a codimension-two surface as~\cite{Dong:2013qoa}
\begin{equation}
\begin{split}
S_{\text{EE}} &= 2\pi\int \mathrm{d}^{d-1} y \sqrt{g} \Bigg\{ -\frac{\partial \mathbf{L}}{\partial R_{\mu\rho\nu\sigma}} \varepsilon_{\mu\rho} \varepsilon_{\nu\sigma} \\
&\qquad + \sum_\alpha \left(\frac{\partial^2 \mathbf{L}}{\partial R_{\mu_1\rho_1\nu_1\sigma_1} \partial R_{\mu_2\rho_2\nu_2\sigma_2}}\right)_\alpha \frac{2K_{\lambda_1\rho_1\sigma_1} K_{\lambda_2\rho_2\sigma_2}}{q_\alpha+1}\times \\
&\qquad \times \left[ (n_{\mu_1\mu_2} n_{\nu_1\nu_2}-\varepsilon_{\mu_1\mu_2} \varepsilon_{\nu_1\nu_2}) n^{\lambda_1\lambda_2} + (n_{\mu_1\mu_2} \varepsilon_{\nu_1\nu_2}+\varepsilon_{\mu_1\mu_2} n_{\nu_1\nu_2}) \varepsilon^{\lambda_1\lambda_2} \right] \Bigg\} \,.
\end{split}
\label{eq:Dong_Covariant}
\end{equation}
Here, $y$ and $g$ represent the coordinates and the determinant of the induced metric on the $(d-1)$-dimensional surface, while $\mathbf{L}$ is the higher-derivative Lagrangian. The tensors $n_{\mu\nu}$, $\varepsilon_{\mu\nu}$, and $K_{\lambda\mu\nu}$ naturally encode the projection, the Levi-Civita tensor, and the extrinsic curvature of the surface, respectively. Within the brackets, the first term represents the standard Wald entropy. In the subsequent anomaly correction term, the summation over $\alpha$ organizes the expansion with an associated anomaly coefficient $q_\alpha$, and the entire third line acts purely as a projector. Noticeably, $W_{\mathrm{gen}}$ mirrors the Wald formula, whereas $W_K$ perfectly parallels Dong's anomaly corrections.

Despite these formal developments, the application of this complete volume functional to both static and dynamical black holes remains entirely unexplored. In this paper, we aim to bridge this gap by systematically investigating holographic complexity in $(d+1)$-dimensional Gauss-Bonnet gravity via the aforementioned complete volume proposal. Our investigation proceeds in three logical stages: First, as a solid foundation, we derive exact analytical expressions for the late-time behavior and provide the full numerical time evolution for unperturbed eternal black holes. Second, building upon this static background, we introduce dynamics by evaluating the one-sided Vaidya spacetime. Here, via a small Gauss-Bonnet coupling expansion, we analytically evaluate the complexity growth rate for the collapsing null shell geometry. Finally, we explore the scrambling of information in two-sided Vaidya spacetimes. We analytically evaluate the early time, late time, and transient plateau limits, carefully differentiating critical times in heavy and light shock limits.

The remainder of this paper is organized as follows. In section~\ref{sec:Eternal_Gauss-Bonnet_Black_Hole}, we briefly review the complete volume functional and present our static results for the unperturbed Gauss-Bonnet black hole. In section~\ref{sec:Vaidya}, we extend our analysis to dynamical settings, detailing the calculations for both the one-sided collapse and the two-sided shock wave geometries. We summarize our findings and discuss future directions in section~\ref{sec:Discussion}.

\section{Eternal Gauss-Bonnet Black Hole}\label{sec:Eternal_Gauss-Bonnet_Black_Hole}

\begin{figure}[htbp]
	\centering
\includegraphics[scale=0.7]{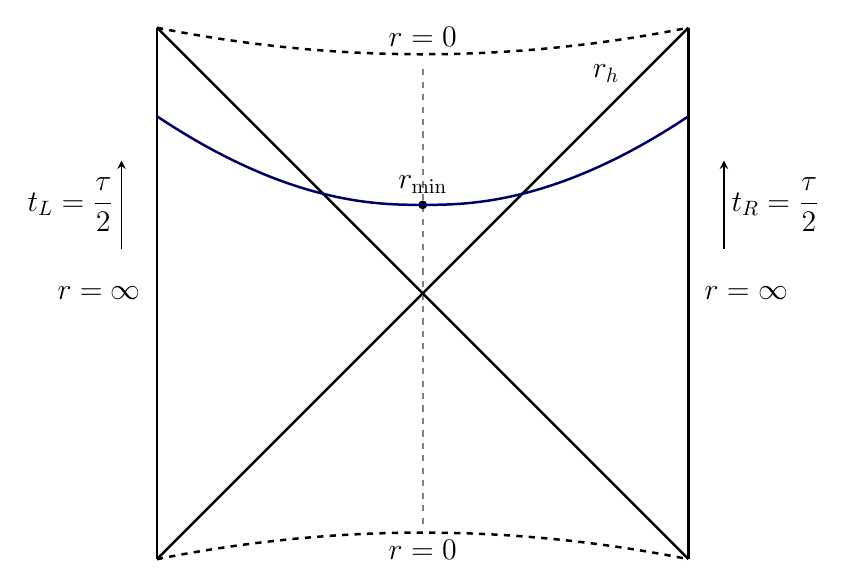}
	\caption{Penrose diagram of the unperturbed eternal Gauss-Bonnet AdS black hole. The dark blue curve represents the extremal codimension-one bulk hypersurface $\mathcal{B}$. It anchors at the left asymptotic boundary ($r=\infty$) at time $\tL$, crosses the future event horizon $r_h$ to reach a minimal radius $r_{\text{min}}$ inside the black hole, and extends to the right boundary at time $\tR$. The diagram illustrates a symmetric time evolution with $\tL = \tR = \tau/2$.}
	\label{fig_eternal_BH}
\end{figure}

The unperturbed eternal black hole geometry is dual to a thermofield double (TFD) state~\cite{Maldacena:2001kr}. This pure state entangles two identical copies of the CFT living on the independent left (L) and right (R) asymptotic boundaries (see figure~\ref{fig_eternal_BH}). Explicitly, the TFD state is constructed as
\begin{equation}
| \Psi_{\text{TFD}} \rangle = \frac{1}{\sqrt{Z}} \sum_{n = 0}^{\infty} e^{-\frac{\beta E_n}{2}} | E_n \rangle_\mt{L} \otimes | E_n \rangle_\mt{R} \,.
\label{eq:TFD_def}
\end{equation}
Under independent time evolution applied to both boundaries via the unitary operators $\UL(\tL) = e^{-i H_\mt{L} \tL}$ and $\UR(\tR) = e^{-i H_\mt{R} \tR}$, the time-dependent TFD state becomes 
\begin{eqnarray}
| \Psi_{\text{TFD}} (\tL, \tR) \rangle &=& \UL (\tL) \, \UR (\tR) \, | \Psi_{\text{TFD}} \rangle \nonumber \\ 
&=& \frac{1}{\sqrt{Z}} \sum_{n = 0}^{\infty} e^{-\frac12\,\beta E_n-iE_n(\tL+\tR)}\, | E_n \rangle_\mt{L} \otimes | E_n \rangle_\mt{R} \,. 
\label{TFDState}
\end{eqnarray}

An inspection of the phase factor in eq.~\eqref{TFDState} reveals a crucial symmetry: the state remains invariant under the time shift 
\begin{equation}
\tL \to \tL + \delta t \,, \qquad \tR \to \tR - \delta t \,. 
\label{eq:time_shift}
\end{equation} 
This implies that the state is invariant if we time evolve with the combined Hamiltonian $H_\mt{L} - H_\mt{R}$, restricting its nontrivial evolution solely to the sum $\tL + \tR$~\cite{Stanford:2014jda,Brown:2015lvg,Carmi:2017jqz}.

\subsection{Generalization of the CV Conjecture to Gauss-Bonnet Gravity}

The action for Gauss-Bonnet gravity is given by
\begin{equation}I_{\mathrm{bulk}}^{\mathrm{GB}}=\frac{1}{16\pi G_{\mathrm{bulk}}}\int \mathrm{d}^{d+1}x\sqrt{-g}\left[\frac{d(d-1)}{L^{2}}+\mathcal{R}[g_{\mu\nu}]+\alpha\mathcal{L}_{\mathrm{GB}}\right]+I_{\mathrm{surf}}^{\mathrm{GB}},\label{Gauss-Bonnet_action}\end{equation}
where
\begin{equation}\alpha=\frac{\tilde{\alpha}}{(d-2)(d-3)},\quad\mathcal{L}_{\mathrm{GB}}=\mathcal{R}_{\mu\nu\rho\sigma}\mathcal{R}^{\mu\nu\rho\sigma}-4\mathcal{R}_{\mu\nu}\mathcal{R}^{\mu\nu}+\mathcal{R}^2.\end{equation}
Here, the cosmological constant is $\Lambda = -d(d-1)/(2L^2)$, $\mathcal{R}[g_{\mu\nu}]$ represents the Ricci scalar, and the surface term $I_{\mathrm{surf}}^{\mathrm{GB}}$ ensures a well-posed variational principle~\cite{Lovelock:1971yv}. The Gauss-Bonnet coupling constant $\alpha$ is positive in heterotic string theory~\cite{Boulware:1985wk}. In particular, for $d = 4$, causality constraints in the boundary CFT impose the bound $-7/72 \leq \alpha/L^2 \leq 9/200$~\cite{Brigante:2007nu,Brigante:2008gz,Buchel:2009tt,Hofman:2009ug}. Accordingly, combining these conditions, we restrict our subsequent numerical analysis to the range $0 \leq \alpha/L^2 \leq 9/200$, which also ensures the validity of treating $\alpha$ as a perturbative parameter.
Substituting the GB Lagrangian into eq.~\reef{eq:W_definitions}, the $W_{\mathrm{gen}}$ term yields contributions involving the bulk curvature components $\mathcal{R}[g_{\mu\nu}]$ and $\mathcal{R}^{\mu\nu}n_{\mu}n_{\nu}$, whereas the $W_K$ term produces combinations of the extrinsic curvature $\mathcal{K}_{\mu\nu}\mathcal{K}^{\mu\nu} - \mathcal{K}^2$. By explicitly employing Gauss's ``Theorema Egregium'' for the hypersurface $\mathcal{B}$, these contributions can be elegantly recast purely in terms of the intrinsic scalar curvature on the hypersurface $\mathcal{B}$
\begin{equation}
R_{\mathcal{B}}[h_{\alpha\beta}] = \mathcal{R}[g_{\mu\nu}] + 2\mathcal{R}^{\mu\nu}n_{\mu}n_{\nu} - \mathcal{K}^2 + \mathcal{K}_{\mu\nu} \mathcal{K}^{\mu\nu} \,.
\label{eq:Theorema_Egregium}
\end{equation}
Consequently, the complete volume functional for GB gravity evaluates to
\begin{align}
  W_{\mathrm{gen}}(\mathcal{B})+W_K(\mathcal{B})&=\int_{\mathcal{B}}\mathrm{d}^{d-1}\sigma \mathrm{d} z \sqrt{\det h}\left(1+\frac{2 \tilde{\alpha}}{(d-1)(d-2)} R_{\mathcal{B}}\right) \,,
  \label{bulk_Vol}\\
  W_{\mathrm{bdy}}(\widetilde{\mathcal{B}})&=\frac{4 \tilde{\alpha}}{(d-1)(d-2)} \int_{\widetilde{\mathcal{B}}} \mathrm{d}^{d-1} \sigma K\sqrt{\det \tilde{h}} \,. 
\end{align}
In this paper, the variational boundary condition simply dictates that the hypersurface $\mathcal{B}$ is anchored at the boundary time slice $\tau$.

To find the extremal hypersurface, we extremize the action incorporating second-order derivatives 
\begin{equation}
W_{\mathrm{bulk}} = W_{\mathrm{gen}}(\mathcal{B}) + W_K(\mathcal{B}) = \int \mathbf{L}_{\mathrm{bulk}}(x^\mu, \dot{x}^\mu, \ddot{x}^\mu)\, \mathrm{d}\lambda \,,
\label{bulk_action}
\end{equation}
where we parametrize the embedding coordinates $x^\mu$ using an intrinsic radial coordinate $\lambda$ and dots denote derivatives with respect to $\lambda$.

The variation of $W_{\mathrm{bulk}}$ yields
\begin{equation}
\begin{aligned}
\delta W_{\mathrm{bulk}} = & 
\left[ P_\mu \delta x^\mu + \Pi_\mu\delta \dot{x}^\mu \right]_{\mathrm{bdy}} \\
& + \int \left[ \frac{\partial \mathbf{L}_{\mathrm{bulk}}}{\partial x^\mu} - \frac{\mathrm{d}}{\mathrm{d}\lambda} \left( \frac{\partial \mathbf{L}_{\mathrm{bulk}}}{\partial \dot{x}^\mu} \right) + \frac{\mathrm{d}^2}{\mathrm{d}\lambda^2} \left( \frac{\partial \mathbf{L}_{\mathrm{bulk}}}{\partial \ddot{x}^\mu} \right) \right] \delta x^\mu\, \mathrm{d}\lambda \,,
\end{aligned}
\label{bulk_variation}
\end{equation}
where $P_\mu$ and $\Pi_\mu$ are the Ostrogradsky momenta defined as
\begin{equation}
  P_\mu \equiv \frac{\partial \mathbf{L}_{\mathrm{bulk}}}{\partial \dot{x}^\mu} - \frac{\mathrm{d}}{\mathrm{d}\lambda}\left( \frac{\partial \mathbf{L}_{\mathrm{bulk}}}{\partial \ddot{x}^\mu} \right) \,, \quad
  \Pi_\mu \equiv \frac{\partial \mathbf{L}_{\mathrm{bulk}}}{\partial \ddot{x}^\mu} \,. 
\end{equation} 
Requiring the bulk variation to vanish yields the equation of motion
\begin{equation}
\frac{\mathrm{d}^2}{\mathrm{d}\lambda^2} \left( \frac{\partial \mathbf{L}_{\mathrm{bulk}}}{\partial \ddot{x}^\mu} \right) - \frac{\mathrm{d}}{\mathrm{d}\lambda} \left( \frac{\partial \mathbf{L}_{\mathrm{bulk}}}{\partial \dot{x}^\mu} \right) + \frac{\partial \mathbf{L}_{\mathrm{bulk}}}{\partial x^\mu} = 0 \,.
\label{eq:EOM_v}
\end{equation}
Since the Lagrangian does not depend explicitly on $v$, we observe from~\reef{eq:EOM_v} that the corresponding conserved momentum is
\begin{equation}
  P_v \equiv \frac{\partial \mathbf{L}_{\mathrm{bulk}}}{\partial \dot{v}} - \frac{\mathrm{d}}{\mathrm{d}\lambda} \left( \frac{\partial \mathbf{L}_{\mathrm{bulk}}}{\partial \ddot{v}} \right).
  \label{Pv}
\end{equation}
For notational simplicity in the subsequent sections, we will frequently denote the conjugate momentum $P_v$ simply as $P$, restoring the subscript only when it is necessary to distinguish it from the momentum $P_r$.
Crucially, the full variational problem contains the usual Gibbons-Hawking-York (GHY) boundary term $W_{\mathrm{bdy}}$. Its variation precisely cancels the higher-derivative contributions 
\begin{equation}
\delta W_{\mathrm{bdy}} = - \left[ \Pi_\mu\delta \dot{x}^\mu \right]_{\mathrm{bdy}} \,. 
\label{eq:variation_bdy}
\end{equation}
Thus, combining eq.~\reef{eq:variation_bdy} and eq.~\reef{bulk_variation} reveals that the complexity growth rate remains governed by the conserved momentum $P$.

\subsubsection*{Black Hole Solution}
The exact line element for the static black hole solution in $(d+1)$ dimensions, derived from the action~\reef{Gauss-Bonnet_action}, is given by~\cite{Boulware:1985wk,Cai:2001dz}
\begin{equation}
  \mathrm{d}s^2 = -f(r)\mathrm{d}t^2 + \frac{\mathrm{d}r^2}{f(r)} +r^2 h_{ij} \mathrm{d}x^i \mathrm{d}x^j \,, 
\label{tr_metric}
\end{equation}
where $h_{ij}\mathrm{d}x^{i}\mathrm{d}x^{j}$ represents the line element of a $(d-1)$-dimensional hypersurface with constant curvature $(d-1)(d-2)k$. 
The corresponding metric function $f(r)$ reads 
\begin{equation}
  f(r) = k + \frac{r^2}{2\tilde{\alpha}} \left(1 \pm \sqrt{1 + \frac{64\pi G_{\mathrm{bulk}}\tilde{\alpha} M}{(d-1)\Omega_{k,d-1} r^d} - \frac{4\tilde{\alpha}}{L^2}}\right) \,.
\end{equation}
Here, $M$ denotes the gravitational mass, and $\Omega_{k,d-1}$ represents the dimensionless volume of the $(d-1)$-dimensional maximally symmetric submanifold. The horizon topology is determined by the parameter $k \in \{1, 0, -1\}$, corresponding to a spherical, planar, or hyperbolic geometry, respectively. The explicit form of the line element $\mathrm{d}\Omega_{k,d-1}^2 \equiv h_{ij}\mathrm{d}x^i \mathrm{d}x^j$ is given by
\begin{equation}
\mathrm{d}\Omega_{k,d-1}^2 =
\begin{cases}
\mathrm{d}\theta^2 + \sin^2\theta \, \mathrm{d}\Omega_{d-2}^2 \,, & k = 1 \,, \\[4pt] 
\sum_{i=1}^{d-1} \mathrm{d}x_i^2 \,, & k = 0 \,, \\[4pt]
\mathrm{d}\theta^2 + \sinh^2\theta \, \mathrm{d}\Omega_{d-2}^2 \,, & k = -1 \,,
\end{cases}
\end{equation}
where $\mathrm{d}\Omega_{d-2}^2$ denotes the metric of the $(d-2)$-dimensional unit sphere.
Following standard conventions in the literature, we trade the gravitational mass $M$ for the mass parameter $\omega$ using the relation
\begin{equation}
  M = \frac{(d-1)\Omega_{k,d-1}}{16\pi G_{\mathrm{bulk}}} \omega^{d-2} \,.
\end{equation}
Consequently, the metric function takes the form
\begin{equation}
  f(r) = k + \frac{r^2}{2\tilde{\alpha}} \left( 1 \pm \sqrt{1 + 4\tilde{\alpha}\frac{\omega^{d-2}}{r^d} - \frac{4\tilde{\alpha}}{L^2}} \right) \,. 
  \label{eq:main_metric}
\end{equation}
To recover the standard Schwarzschild-AdS geometry in the Einstein gravity limit ($\tilde{\alpha} \to 0$), we must select the negative branch of the solution, which smoothly reduces to
\beq
f(r) \to k - \frac{\omega^{d-2}}{r^{d-2}} + \frac{r^2}{L^2} \,.
\eeq
Conversely, the positive branch is typically discarded as it exhibits ghost instabilities and lacks a physically well-defined vacuum state~\cite{Cai:2001dz}.
The radius of the event horizon, $r_h$, is determined by $f(r_h)=0$. This condition allows us to express the mass parameter $\omega$ in terms of the horizon radius as
\beq
\omega^{d-2} = \tilde{\alpha} k^2 r_h^{d-4} + k r_h^{d-2} + \frac{r_h^d}{L^2} \,. 
\label{M_omega_relation}
\eeq
For the static background metric~\reef{tr_metric}, the Euclidean section must be free of conical singularities at the horizon. This regularity condition fixes the Hawking temperature as $T = 1/\beta = (4\pi)^{-1} \partial_r f \big|_{r=r_h}$. Together with the black hole entropy $S$, obtained via the first law of thermodynamics, these physical quantities are given by~\cite{Cai:2013qga}
\begin{equation}
\begin{aligned}
T &= \frac{d\, r_h^4 + (d-2)k L^2 r_h^2 + \tilde{\alpha} (d-4)k^2 L^2}{4\pi L^2 r_h (2\tilde{\alpha} k + r_h^2)}\,, \\[4pt]
S &= \frac{\Omega_{k,d-1}\, r_h^{\,d-1}}{4G_{\mathrm{bulk}}} \left(1 + \frac{d-1}{d-3}\,\frac{2\tilde{\alpha} k}{r_h^2}\right) \,. 
\end{aligned}
\label{temperature_entropy}
\end{equation}
To facilitate the subsequent calculations, we introduce the following dimensionless variables
\begin{equation}
  x \equiv \frac{r}{r_h} \,, \qquad 
  \hat{\alpha} \equiv \frac{\tilde{\alpha}}{L^2} \,, \qquad 
  \epsilon \equiv \frac{L}{r_h} \,.
\label{eq:dimensionless_variables}
\end{equation}
These variables allow us to conveniently express bulk physical quantities in terms of boundary variables.
By expanding the temperature relation to first order in $\hat{\alpha}$, we can express $\epsilon$ as a function of the dimensionless quantity $LT$
\begin{equation}
  \epsilon = \frac{d}{4\pi L T} + \frac{d^2 k (d - 2 - 2d\hat{\alpha})}{(4\pi L T)^3} + \cdots \,.
\label{eq:zfuncT}
\end{equation}
Notably, for black holes with a planar horizon topology ($k=0$), all higher-order corrections identically vanish, yielding the exact relation
\begin{equation}
  \epsilon = \frac{d}{4\pi L T} \,.
\label{eq:zfuncT_planar}
\end{equation}

\subsubsection*{Equations of Motion for the Complete Volume Functional}

To evaluate the complete volume functional, the first step is to rewrite the static metric~\eqref{tr_metric} in Eddington-Finkelstein coordinates
\begin{equation}
\mathrm{d}s^2 = -f(r)\mathrm{d}v^2 + 2\,\mathrm{d}v\,\mathrm{d}r + r^2\mathrm{d}\Omega_{k,d-1}^2 \,. 
\label{eq:Eddington_Finkelstein_coordinates}
\end{equation}

Evaluating the functional~\eqref{bulk_Vol} requires the explicit intrinsic scalar curvature $R_{\mathcal{B}}$ of the spacelike hypersurface $\mathcal{B}$. As previously established, we parameterize this codimension-one hypersurface $\mathcal{B}$ as $\{v(\lambda), r(\lambda)\}$. The induced metric on $\mathcal{B}$ is then
\begin{equation}
\mathrm{d}s^2_{\mathcal{B}} = \bigl[-f(r) \dot{v}^2 + 2\dot{v}\dot{r}\bigr] \mathrm{d}\lambda^2 + r^2 \mathrm{d}\Omega_{k,d-1}^2 \,.
\end{equation}

Recall that a generic $D$-dimensional total manifold $\mathcal{P} = \mathcal{M} \times \mathcal{F}$ (with $D = m + n$) can be equipped with a warped-product metric
\begin{equation}
\mathrm{d}s^2 = g_{\mu\nu} \mathrm{d}x^\mu \mathrm{d}x^\nu = h_{ab}(y) \mathrm{d}y^a \mathrm{d}y^b + R^2(y) \gamma_{\alpha\beta}(z) \mathrm{d}z^\alpha \mathrm{d}z^\beta \,, 
\end{equation}
where $y^a$ ($a=1,\dots,m$) and $z^\alpha$ ($\alpha=1,\dots,n$) parameterize the base manifold $\mathcal{M}$ and the fiber manifold $\mathcal{F}$, respectively. Here, $h_{ab}$ and $\gamma_{\alpha\beta}$ denote their corresponding intrinsic metrics, and the positive function $R(y)$ acts as the warp factor.

The scalar curvature of this total manifold evaluates to
\begin{equation}
R^{(\mathrm{total})} = R^{(h)} + \frac{R^{(\gamma)}}{R^2(y)} - 2n \frac{\Box_h R(y)}{R(y)} - n(n-1) \frac{(\nabla^{(h)} R(y))^2}{R^2(y)} \,, 
\label{eq:total_Ricci}
\end{equation}
where $R^{(h)}$ and $R^{(\gamma)}$ denote the intrinsic Ricci scalars of the base and the fiber. The differential operators on the base manifold are explicitly defined as $\Box_h R(y) \equiv h^{ab} \nabla^{(h)}_a \nabla^{(h)}_b R(y)$ and $(\nabla^{(h)} R(y))^2 \equiv h^{ab} \big(\nabla^{(h)}_a R(y)\big)\big(\nabla^{(h)}_b R(y)\big)$.

Applying eq.~\reef{eq:total_Ricci} to our specific induced metric, a direct computation yields the scalar curvature $R_{\mathcal{B}}$ of the spacelike hypersurface $\mathcal{B}$:
\begin{equation}
\begin{aligned}
R_{\mathcal{B}} &= \frac{d-1}{r^2 \bigl(2\dot{r}\dot{v} - f(r)\dot{v}^2\bigr)^2} \\
&\quad \times \Bigl[ \bigl(2\dot{r}\dot{v} - f(r)\dot{v}^2\bigr) \bigl((d-2)\bigl(k(2\dot{r}\dot{v} - f(r)\dot{v}^2) - \dot{r}^2\bigr) - 2r\ddot{r}\bigr) \\
&\quad \quad + r\dot{r}\bigl(-\dot{r}\dot{v}^2 f'(r) - 2f(r)\dot{v}\ddot{v} + 2\ddot{r}\dot{v} + 2\dot{r}\ddot{v}\bigr) \Bigr] \,.
\end{aligned}
\label{eq:R_B_explicit}
\end{equation}
The conserved momentum obtained from eq.~\reef{Pv} then evaluates to
\begin{equation} 
P = \frac{r^{d-3} \left(\dot{r}-f(r) \dot{v}\right) \left[f(r) \dot{v}^2 \left(r^2+2 \tilde{\alpha} k\right)+2 \tilde{\alpha} \dot{r}^2-2 \dot{r} \dot{v} \left(r^2+2 \tilde{\alpha} k\right)\right]}{\dot{v} \sqrt{\dot{v} \left(2 \dot{r}-f(r) \dot{v}\right)} \left(f(r) \dot{v}-2 \dot{r}\right)} \,. 
\label{eq:conserved_momentum}
\end{equation}
Owing to the reparameterization invariance of the Lagrangian, we can freely choose $\lambda$ to fix the radial volume element. Here we adopt the gauge condition
\begin{equation}
  r^{d-1} = \sqrt{-f(r) \dot{v}^2 + 2 \dot{v} \dot{r}} \,. 
  \label{eq:gauge_constraint}
\end{equation}
Solving the gauge constraint~\eqref{eq:gauge_constraint} for $\dot{v}$ yields
\begin{equation}
\dot{v} = \frac{\dot{r} \mp \sqrt{\dot{r}^2 - r^{2d-2} f(r)}}{f(r)} \,. 
\end{equation}
Substituting this expression back into the conserved momentum $P$ to eliminate $\dot{v}$, we obtain
\begin{equation} 
P = \pm r^{-2d-2} \sqrt{\dot{r}^2-r^{2 d-2} f(r)} \left(r^{2 d} \left(r^2+2 \tilde{\alpha} k\right)-2 \tilde{\alpha} r^2 \dot{r}^2\right) \,. 
\label{eq:Conserved_momentum} 
\end{equation} 
Unlike standard particle motion where the effective potential depends solely on $r$ and the conserved quantity acts as an additive constant, the highly non-linear relation in eq.~\eqref{eq:Conserved_momentum} prevents such a separation. Therefore, we define the effective potential directly as $\mathcal{U}(P, r)\equiv-\dot{r}^2$.
Calculation shows that, regardless of which branch of $P$ is chosen, one obtains a unified expression for $\dot{v}$ as a function of $\dot{r}$:
\beq
\dot{v}=\frac{1}{f(r)}\left(\frac{P}{2 \tilde{\alpha} r^{-2 d} \dot{r}^2-\frac{2 \tilde{\alpha} k}{r^2}-1}+\dot{r}\right) \,. 
\label{dot_v}
\eeq
Squaring the momentum relation~\eqref{eq:Conserved_momentum} to solve for $X \equiv \dot{r}^2 = -\mathcal{U}$, we find that the higher-curvature corrections result in a \textit{cubic} equation for $X$
\begin{equation}
  \mathcal{C}_3 X^3 + \mathcal{C}_2 X^2 + \mathcal{C}_1 X + \mathcal{C}_0 = 0 \,,
  \label{eq:cubic_EOM}
\end{equation}
with the coefficients reading
\begin{align}
  \mathcal{C}_3 &= 4 \tilde{\alpha}^2 r^{-2d-2} \,, \\
  \mathcal{C}_2 &= - 4\tilde{\alpha} r^{-4} \bigl[ \tilde{\alpha} f(r) + 2\tilde{\alpha} k + r^2 \bigr] \,, \\
  \mathcal{C}_1 &= r^{2d-6} (2\tilde{\alpha} k + r^2) \bigl[ 4\tilde{\alpha} f(r) + 2\tilde{\alpha} k + r^2 \bigr] \,, \\
  \mathcal{C}_0 &= -r^{4d-8} f(r) (2\tilde{\alpha} k + r^2)^2 - P^2 r^{2d-2} \,.
\end{align}

While exact analytical roots for $\dot{r}^2$ can be derived via Cardano's formula (presented in appendix~\ref{appendix:A} for completeness), their cumbersome algebraic forms offer limited physical insight. Instead, examining the Einstein gravity limit ($\tilde{\alpha} \to 0$) proves highly instructive. In this limit, the coefficients $\mathcal{C}_3$ and $\mathcal{C}_2$ vanish, reducing the cubic equation to a linear one. Consequently, only one solution branch remains finite, smoothly recovering the standard Einstein result $\dot{r}^2 = P^2 + r^{2d-2}f(r)$. The remaining two roots represent non-perturbative branches that exhibit singular asymptotic behavior, diverging as $\mathcal{O}(1/\tilde{\alpha})$.

\begin{figure}[htbp]
	\centering
  \includegraphics[scale=0.7]{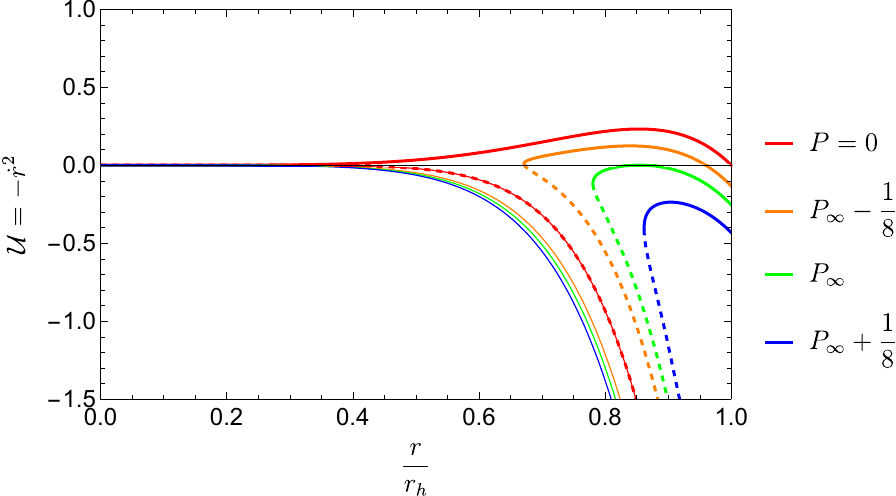} 
	\caption{
The effective potential $\mathcal{U}(P, r) = -\dot{r}^2$ for the planar Gauss-Bonnet black hole in $d=4$ (for illustrative purposes, we set $\alpha/L^2 = 9/200$) for various values of the conserved momentum $P$. The colors distinguish different momentum values, including the initial hypersurface with $P=0$ (red) and the final slice with $P = P_{\infty}$ (green). For each $P$, the cubic equation of motion admits three distinct branches: The thick solid lines represent the Einstein-like branch, the sole physical branch. 
  The dashed and thin solid lines correspond to the non-perturbative branches. These two branches either terminate strictly at the singularity or fail to exhibit any turning point. Furthermore, these two branches overlap at $P=0$.}
\label{fig_effective_potential_planar_a}
\end{figure}

For the numerical results presented below, we express the Gauss-Bonnet coupling in terms of the dimensionless parameter $\alpha/L^2$, which relates to our analytical parameter in $d=4$ via $\hat{\alpha} = 2(\alpha/L^2)$. To visualize these solutions, we plot the effective potential for the planar topology ($k=0$) in figure~\ref{fig_effective_potential_planar_a}, while other geometries exhibit similar behavior. We include the non-perturbative branches solely to illustrate the complete algebraic structure of the cubic equation.

Crucially, the extremal surfaces we are interested in here require the existence of a smooth turning point $r_{\mathrm{min}}$ with $\dot{r} \big|_{r=r_{\mathrm{min}}}=0$ (where the hypersurface reaches its minimal radial position), allowing the surface to bridge the left and right asymptotic boundaries symmetrically. Analyzing $\dot{r}_i^2$ (with $i=1,2,3$) in appendix~\ref{appendix:A} reveals the distinct nature of each root. 
(i) $\dot{r}_1^2$ exhibits $\dot{r}^2 > 0$ globally. Lacking a turning point, it represents a disconnected surface that fails to connect the two asymptotic boundaries.
(ii) $\dot{r}_2^2$, the ``Einstein-like'' branch, possesses a valid and momentum-dependent turning point. It smoothly bridges the two asymptotic boundaries and correctly recovers the Einstein gravity limit as $\tilde{\alpha} \to 0$. 
(iii) $\dot{r}_3^2$ satisfies the turning condition $\dot{r}=0$ only at the singularity $r=0$, thus failing to form a smooth Einstein-Rosen bridge.

With our hypersurface $\mathcal{B}$ uniquely specified by this Einstein-like branch and determined by eq.~\reef{dot_v} and eq.~\eqref{eq:Y2} in appendix~\ref{appendix:A}, we proceed with our computation following the standard framework of~\cite{Carmi:2017jqz}, fixing the arbitrary length scale in the complexity functional to the AdS radius by setting $\ell = L$. By setting $\dot{r}=0$ in eq.~\eqref{eq:Conserved_momentum}, we determine the minimal radius by solving
\begin{equation}
P = r_{\mathrm{min}}^{d-3} \sqrt{-f(r_{\mathrm{min}})}\bigl(r_{\mathrm{min}}^2 + 2\tilde{\alpha} k\bigr) \,.
\label{eq:Pv_rmin}
\end{equation}
This condition is equivalent to $\mathcal{U}(P, r_{\mathrm{min}}) = 0$.

Geometrically, $P = 0$ corresponds to the initial time slice where the turning point touches the event horizon. Conversely, for momenta exceeding a critical value $P_{\infty}$, the turning point vanishes, implying that the surface fails to bridge the two boundaries. Valid two-sided extremal surfaces thus exist exclusively within the momentum window
\begin{equation}
  0 \leq P \leq P_{\infty} \,. 
\end{equation}

\begin{figure}[htbp]
	\centering
	\includegraphics[scale=0.7]{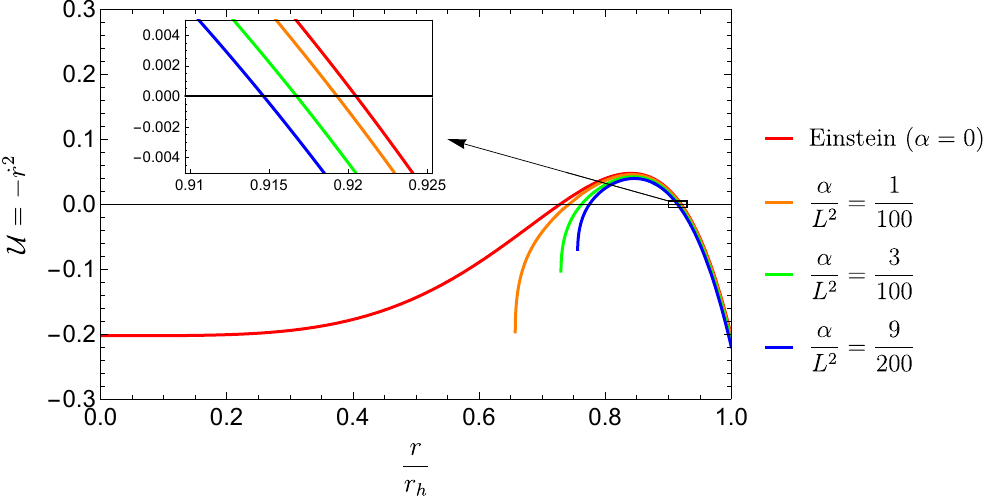}
\caption{
  The effective potential $\mathcal{U}(P, r) = -\dot{r}^2$ of the extremal surface as a function of the normalized radius $r/r_h$ in the $d=4$ planar case ($k=0$). 
  The conserved momentum is fixed at an intermediate value $P = (9/10) P_{\mathrm{crit}}^{(0)}$, where $P_{\mathrm{crit}}^{(0)}$ is the critical momentum in the Einstein limit.
  The curves correspond to standard Einstein gravity ($\alpha/L^2=0$, red) and Gauss-Bonnet gravity with dimensionless couplings $\alpha/L^2 = 1/100$ (orange), $3/100$ (green), and $9/200$ (blue). 
  The inset shows that for a fixed momentum $P$, a larger coupling $\alpha/L^2$ shifts the turning point towards the singularity, implying that the extremal surface penetrates deeper into the bulk. 
}
	\label{fig_effective_potential_alpha_planar_a}
\end{figure}

To explicitly separate the stringy effects from the boundary time evolution, we temporarily treat $P$ as a fixed parameter to explore the deformed phase space of the extremal slices. Taking the planar topology ($k=0$) as a representative example, figure~\ref{fig_effective_potential_alpha_planar_a} illustrates the effective potential of this physical branch under varying Gauss-Bonnet couplings $\alpha/L^2$. As highlighted in the inset, for a fixed conserved momentum, increasing the positive coupling $\alpha/L^2$ monotonically shifts the zero of the potential (the turning point $r_{\mathrm{min}}$) closer to the singularity. This indicates that the higher-derivative corrections cause the extremal surface to penetrate deeper into the black hole interior compared to standard Einstein gravity.

Remarkably, even for the generalized volume functional, the complexity growth rate remains strictly governed by the conserved momentum $P$. Consequently, the rate of complexity growth is compactly expressed as
\begin{equation}
  \frac{\mathrm{d}\cv}{\mathrm{d}\tau}= \frac{\Omega_{k,d-1}}{G_{\mathrm{bulk}} L}\,P(\tau) 
  = \pm\frac{\Omega_{k,d-1}}{G_{\mathrm{bulk}} L}\, r_{\mathrm{min}}^{d-3}\sqrt{-f(r_{\mathrm{min}})}\;\bigl(r_{\mathrm{min}}^2 + 2\tilde{\alpha} k\bigr) \,. 
\label{eq:growth_rate} 
\end{equation}

\subsection{Late-Time Behavior}
\label{sec:Late_Time_Behavior}
To analyze the late-time complexity growth, we define the function $W(r)$ from~\eqref{eq:growth_rate} as
\begin{equation}
  W(r) \equiv r^{d-3} \sqrt{-f(r)} \left(r^2 + 2\tilde{\alpha} k\right) \,.
\end{equation}
The turning point $r_{\mathrm{min}}$ is then determined by
\begin{equation}
  r_{\mathrm{min}}^{2d-6} f(r_{\mathrm{min}}) \left(r_{\mathrm{min}}^2 + 2\tilde{\alpha} k\right)^2 + P^2 = 0 \,.
\end{equation}
This equation generally admits two real roots inside the horizon. The turning point is identified as the outermost root, which characterizes the maximal bulk penetration.

In the late-time limit $\tau \to \infty$, the conserved momentum approaches its critical maximum $P_{\infty}$, and the turning point settles at the local maximum of $W(r)$, denoted $\tilde{r}_{\mathrm{min}}$. This critical radius satisfies $W'(\tilde{r}_{\mathrm{min}}) = 0$, yielding
\begin{equation}
\begin{split}
  W'(\tilde{r}_{\mathrm{min}}) &= -\frac{\tilde{r}_{\mathrm{min}}^{d-4}}{2 \sqrt{-f(\tilde{r}_{\mathrm{min}})}} \biggl[ 2f(\tilde{r}_{\mathrm{min}}) \bigl( (d-1)\tilde{r}_{\mathrm{min}}^2 + 2\tilde{\alpha}(d-3)k \bigr) \\
  &\quad + \tilde{r}_{\mathrm{min}} f'(\tilde{r}_{\mathrm{min}}) \bigl(\tilde{r}_{\mathrm{min}}^2 + 2\tilde{\alpha}k\bigr) \biggr] = 0 \,.
\end{split}
\label{eq:late_time_condition}
\end{equation}
To extract the asymptotic behavior, we expand the growth rate around $\tilde{r}_{\mathrm{min}}$, yielding
\begin{equation}
  \frac{\mathrm{d}\cv}{\mathrm{d}\tau} = \frac{\Omega_{k,d-1}}{G_{\mathrm{bulk}} L} \Bigl[ W(\tilde{r}_{\mathrm{min}}) 
  + \frac{1}{2} W''(\tilde{r}_{\mathrm{min}}) (r_{\mathrm{min}}-\tilde{r}_{\mathrm{min}})^2 + \mathcal{O}\bigl((r_{\mathrm{min}}-\tilde{r}_{\mathrm{min}})^3\bigr) \Bigr] \,.
  \label{eq:latetime_dcv_dt}
\end{equation}
In terms of the effective potential, this corresponds to $\partial_r\mathcal{U}(P_{\infty}, \tilde{r}_{\mathrm{min}}) = 0$. Taking the late-time limit, $r_{\mathrm{min}}$ converges to $\tilde{r}_{\mathrm{min}}$. The complexity growth rate thus saturates to a constant value, confirming the linear growth of complexity
\begin{equation}
  \lim_{\tau\rightarrow\infty} \frac{\mathrm{d}\cv}{\mathrm{d}\tau} = \frac{\Omega_{k,d-1}}{G_{\mathrm{bulk}} L} W(\tilde{r}_{\mathrm{min}}) 
  = \frac{\Omega_{k,d-1}}{G_{\mathrm{bulk}} L}\, \tilde{r}_{\mathrm{min}}^{d-3} \sqrt{-f(\tilde{r}_{\mathrm{min}})}\; \bigl(\tilde{r}_{\mathrm{min}}^2 + 2\tilde{\alpha} k\bigr) \,.
  \label{eq:const_dv_dt}
\end{equation}
For planar black holes ($k=0$), eq.~\eqref{eq:const_dv_dt} shows that late-time corrections to the CV proposal vanish. This is physically expected since the Ricci scalar $R_{\mathcal B}$ vanishes on the final slice. Nevertheless, the two proposals remain distinct, with their differences manifesting during the full time evolution (see section~\ref{sec:General_Time_Dependence}).

To analytically resolve the location of the final slice $\tilde{r}_{\mathrm{min}}$, taking the large black hole limit ($r_h \gg L$) proves highly tractable. The condition~\eqref{eq:late_time_condition} dictates that its numerator must vanish, yielding the simplified condition
\begin{equation}
  2f(\tilde{r}_{\mathrm{min}}) \bigl[ (d-1)\tilde{r}_{\mathrm{min}}^2 + 2\tilde{\alpha}(d-3)k \bigr] + \tilde{r}_{\mathrm{min}} f'(\tilde{r}_{\mathrm{min}}) \bigl(\tilde{r}_{\mathrm{min}}^2 + 2\tilde{\alpha}k\bigr) = 0 \,. 
\label{eq:late_time_condition1}
\end{equation}
To facilitate the perturbative expansion, it is highly convenient to recast the Gauss-Bonnet metric function into the alternative explicit form
\begin{equation}
  f(r) = k + \frac{r^{2}}{2\tilde{\alpha} L} \Biggl( L - \sqrt{L^{2} - 4\tilde{\alpha} + 4\tilde{\alpha} r^{-d} r_h^{d-4} \bigl(\tilde{\alpha} k^{2}L^{2} + k L^{2}r_h^{2} + r_h^{4}\bigr)} \, \Biggr) \,.
  \label{eq:rh_metric}
\end{equation}

Employing the dimensionless variables $\epsilon$ and $x$ defined in eq.~\reef{eq:dimensionless_variables}, we substitute eq.~\reef{eq:rh_metric} into the condition~\eqref{eq:late_time_condition1}. Stripping away the overall non-vanishing factors, the determining equation for the final slice naturally organizes into a perturbative series
\begin{equation}
  \mathcal{C}_0(x) + k \epsilon^2 \mathcal{C}_1(x) + k^2 \epsilon^4 \mathcal{C}_2(x) + \mathcal{O}(\epsilon^6) = 0 \,.
\label{eq:master_expansion}
\end{equation}
The exact coefficient functions evaluate to
\begin{subequations}
\begin{align}
  \mathcal{C}_0(x) &= (2\hat{\alpha}+2)x^d - (2\hat{\alpha} + 1) \,, \\
  \mathcal{C}_1(x) &= -\frac{1}{d x^2} \Big[ \bigl(2(4\hat{\alpha}+1) - 2d(2\hat{\alpha}+1)\bigr)x^d + d(2\hat{\alpha}+1)x^2 - 8\hat{\alpha} + 2d\hat{\alpha} \Big] \,, \\
  \mathcal{C}_2(x) &= -\frac{\hat{\alpha}}{d} \Big[ d(x^2+2) - 8 - 4(d-3)x^{d-2} \Big] x^{-2} \,.
\end{align}
\end{subequations}
The leading-order equation $\mathcal{C}_0(x)=0$ immediately yields the solution
\begin{equation}
  x_0 = \left(\frac{2\hat{\alpha}+1}{2\hat{\alpha}+2}\right)^{1/d} \,.
\end{equation}
Solving eq.~\eqref{eq:master_expansion} order by order with the series $x = x_0 (1 + k \epsilon^2 \delta_1 + k^2 \epsilon^4 \delta_2)$, we obtain the explicit expression for $\tilde{r}_{\mathrm{min}}$:
\begin{equation}
\begin{split}
\tilde{r}_{\mathrm{min}} = x_0 r_h \bigg[ & 1 - \frac{2^{2/d}(d-1)-d}{d^2} \frac{L^2}{r_h^2} k \\ 
& + \frac{(d-1) \left(-d^2 + 2^{\frac{2}{d}+1} d + 2^{4/d} (d-3)(d-1)\right)}{2d^4} \frac{L^4}{r_h^4} k^2 + \mathcal{O}\left(\frac{L^6}{r_h^6}\right) \bigg] \,.
\end{split}
\label{eq:rf_expansion}
\end{equation}
Remarkably, the coupling $\hat{\alpha}$ completely decouples from the topological curvature corrections at this order. All $\hat{\alpha}$-dependence is neatly factorized into the simple overall prefactor $x_0$, which smoothly recovers the standard Einstein gravity limit $\tilde{r}_{\mathrm{min}} \to 2^{-1/d}\, r_h$.

Substituting the location of the constant-$r$ slice~\eqref{eq:rf_expansion} back into the growth rate formula~\eqref{eq:const_dv_dt}, we obtain the analytical late-time complexity growth. It proves insightful to cast this rate into two complementary perturbative series: a large black hole expansion in $L/r_h$ and a high-temperature expansion in $(L T)^{-1}$, yielding
\begin{subequations}
\begin{align}
\lim_{\tau \to \infty} \frac{d-1}{8\pi M} \frac{\mathrm{d}\cv}{\mathrm{d}\tau} 
&= \left(1 - \frac{M_{\mathrm{min}}}{M} \delta_{k,-1} \right) \nonumber \\ 
&\quad \times \left[ \left(1 - \frac{\hat{\alpha}}{2} + \mathcal{O}(\hat{\alpha}^2)\right) + \mathcal{A}_2 \left( \frac{L}{r_h} \right)^2 + \mathcal{A}_4 \left( \frac{L}{r_h} \right)^4 + \dots \right] \,, \label{eq:expansion_geometric} \\
&= \left[ 1 + \mathcal{B}_d (4\pi L T)^{-d} \delta_{k,-1} + \dots \right] \nonumber \\
&\quad \times \left[ \left(1 -\frac{\hat{\alpha}}{2} + \mathcal{O}(\hat{\alpha}^2)\right) + \mathcal{C}_2 (4\pi L T)^{-2} + \mathcal{C}_4 (4\pi L T)^{-4} + \dots \right] \,. \label{eq:expansion_of_late_time_limit2}
\end{align}
\end{subequations}
The first factor on the right-hand side of eq.~\eqref{eq:expansion_of_late_time_limit2} is non-trivial strictly for hyperbolic horizons ($k=-1$). This accounts for the negative minimal mass ($M_{\mathrm{min}}<0$) in this geometry, which shifts the reference mass for complexity growth to $M-M_{\mathrm{min}}$.

The second factor in eq.~\eqref{eq:expansion_of_late_time_limit2} encodes effects from both the coupling $\hat{\alpha}$ and the curvature $k$, with the latter appearing only at second and higher orders. In the high-temperature limit ($LT \gg 1$ and where $k$ becomes negligible), the universal leading term $1-\hat{\alpha}/2$ characterizes the suppression of the growth rate relative to Einstein gravity and recovers the expansion of the uncorrected CV proposal for planar GB black holes ($k=0$)~\cite{An:2018dbz}. Additionally, the expansion in eq.~\eqref{eq:rf_expansion} reveals that this limit is governed by the pre-factor $[(2 \hat{\alpha}+1)/(2 \hat{\alpha}+2)]^{1/d} r_h$. Since this factor grows monotonically with $\hat{\alpha}$, a positive coupling repels the final slice position $\tilde{r}_{\mathrm{min}}$ toward the horizon.

Finally, taking the Einstein gravity limit ($\hat{\alpha} \to 0$), all coefficients ($\mathcal{A}_i$, $\mathcal{B}_d$ and $\mathcal{C}_i$) in eqs.~\eqref{eq:expansion_geometric} and~\eqref{eq:expansion_of_late_time_limit2} accurately reproduce the standard results~\cite{Carmi:2017jqz}, validating our perturbative framework. Explicit expressions for the remaining coefficients are relegated to appendix~\ref{app:coefficients}. Although the universal suppression factor ($- \hat{\alpha}/2$) dominates at large $r_h$ (high temperatures), the terms $\mathcal{C}_2$ and $\mathcal{B}_d$ reveal a nontrivial interplay between curvature and stringy corrections for smaller black holes. Specifically, for spherical black holes ($k=1$), $\mathcal{C}_2$ introduces a positive $\hat{\alpha}$ correction that overcomes the baseline suppression, thereby leading to the anomalous accelerated growth rate at intermediate temperatures. Conversely, for hyperbolic black holes ($k=-1$), the curvature correction flips sign and combines with $\mathcal{B}_d$ to reinforce the suppression across all physical scales. We will see that this analytical understanding fully coincides with our numerical results, matching both the explicitly evaluated late-time growth rate and the asymptotic limit of the full time evolution.
\subsubsection*{Numerical Analysis}
We now numerically evaluate the late-time complexity growth rate in $d=4$ across various horizon radii $r_h/L$ to corroborate our previous analytical expansions. For clarity, we establish a unified graphical convention. We use $\cv$ to denote the complexity, encompassing all volume-related definitions discussed. We compare three distinct scenarios: (i) the standard Einstein gravity $\cv^{(\mathrm{E})}$ ($\alpha/L^2=0$, solid black lines); (ii) the uncorrected geometric volume $\cv^{(\mathrm{GB})}$, i.e., the uncorrected CV proposal evaluated in the Gauss-Bonnet background without additional corrections ($\alpha/L^2 = 9/200$, gray dashed lines), with the dashed style explicitly emphasizing its physical incompleteness; and (iii) our complete and corrected volume functional $\cvcorr$, plotted as solid green and blue lines for representative couplings $\alpha/L^2 = 3/100$ and $9/200$, respectively.

\begin{figure}[htbp]
\centering
\includegraphics[width=0.48\textwidth]{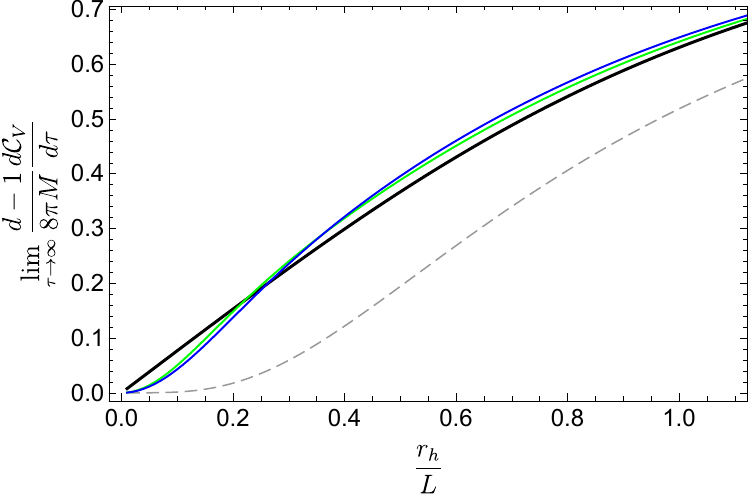}
\hfill \includegraphics[width=0.48\textwidth]{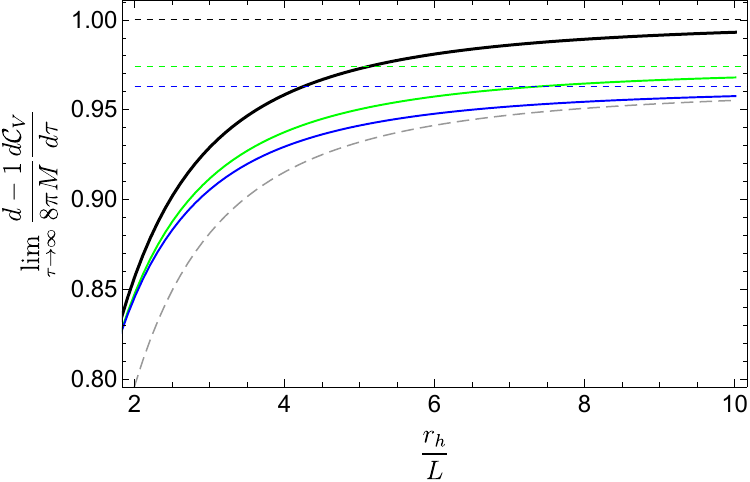}
\caption{Late-time complexity growth rate versus the horizon radius $r_h/L$ for spherical ($k=1$) black holes in $d=4$. The solid black, gray dashed, green, and blue curves correspond to pure Einstein gravity ($\alpha/L^2=0$), the uncorrected GB volume ($\alpha/L^2=9/200$), and the corrected GB volumes with $\alpha/L^2=3/100$ and $9/200$, respectively. Left: $0< r_h/L < 11/10$ case. Right: $r_h/L \geq 2$ case, illustrating the monotonic increase toward the planar asymptotic values (horizontal dashed lines).}
\label{fig_Pvf_rh_Vcorrected_spherical_a1}
\end{figure}

For curved horizons ($k = \pm 1$), the black hole is determined not only by the parameter $\alpha/L^2$ but also by $r_h/L$. Figure~\ref{fig_Pvf_rh_Vcorrected_spherical_a1} illustrates the spherical case ($k=1$), where the rate vanishes at $r_h/L = 0$ and grows monotonically. Interestingly, relative to Einstein gravity, a positive coupling $\alpha/L^2$ suppresses the rate for very small black holes ($r_h/L \to 0$), but enhances it in the intermediate regime (left panel). This non-trivial crossover is captured by the complete volume functional, in stark contrast to the uncorrected CV proposal, which predicts uniform suppression across all $r_h$. For large black holes, the coupling reverts to suppressing the rate, approaching its planar asymptotic limit from below. As indicated by eq.~\reef{eq:expansion_geometric}, topological effects diminish in the large black hole regime ($r_h \gg L$). Consequently, all corrected curves lie entirely below the Einstein gravity, shifting downward as $\alpha/L^2$ increases (right panel).

\begin{figure}[htbp]
\centering
\includegraphics[width=0.48\textwidth]{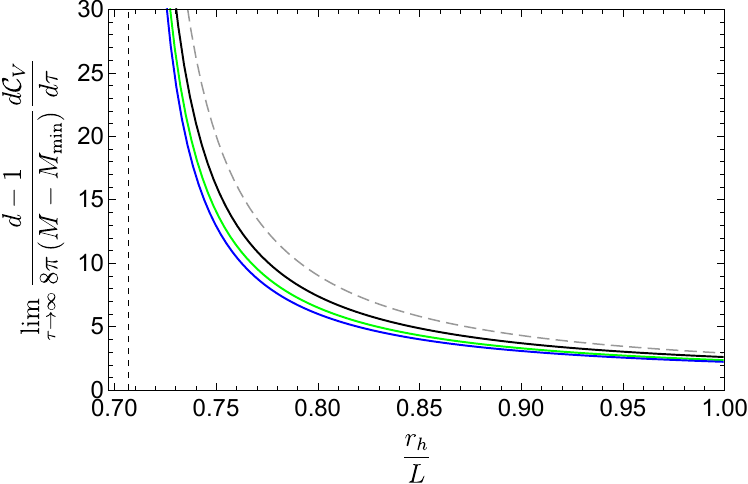}
\hfill \includegraphics[width=0.48\textwidth]{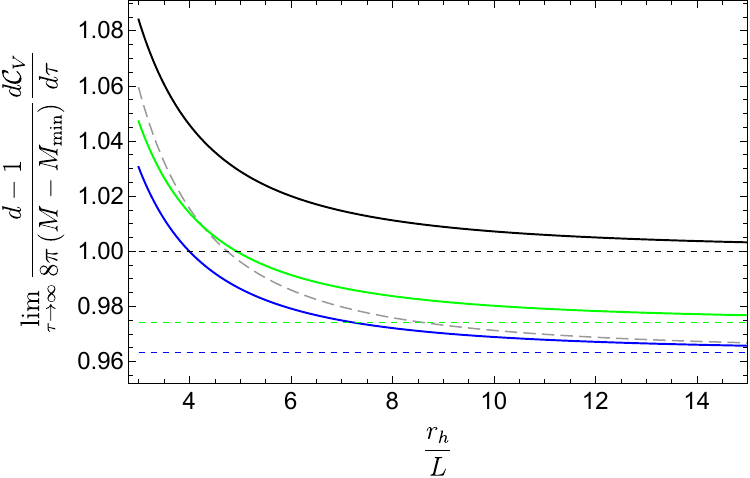}
\caption{Late-time complexity growth rate versus $r_h/L$ for hyperbolic ($k=-1$) black holes in $d=4$. The solid black, gray dashed, green, and blue curves correspond to pure Einstein gravity ($\alpha/L^2=0$), the uncorrected GB volume ($\alpha/L^2=9/200$), and the corrected GB volumes with $\alpha/L^2=3/100$ and $9/200$, respectively. Left: $r_{\mathrm{ext}} \leq r_h \leq L$ case, where vertical lines mark the minimal value $r_{\mathrm{ext}}$ for $r_h$. Right: $r_h/L \geq 3$ case, illustrating the monotonic decrease toward the planar asymptotic values (horizontal dashed lines).}
\label{fig_Pvf_rh_Vcorrected_hyperbolic_a1b1}
\end{figure}

Figure~\ref{fig_Pvf_rh_Vcorrected_hyperbolic_a1b1} illustrates the hyperbolic case ($k=-1$). Here, the range of $r_h/L$ is bounded below by the minimal value $r_{\mathrm{ext}}$ for $r_h$. Unlike the spherical case, the complexity growth rate decreases monotonically and approaches the planar ($k=0$) limit from above. This behavior clearly indicates a violation of Lloyd's bound. The dependence on $k$ gradually disappears as $r_h/L$ grows, similar to the $k=1$ case (right panel). As anticipated from the analytical expansion, the curves uniformly lie below the Einstein gravity, shifting downward with increasing $\alpha/L^2$. Crucially, this suppression persists even in the regime $r_{\mathrm{ext}} \leq r_h \leq L$ (left panel), but the uncorrected geometric volume deviates from this trend and the rate exceeds the Einstein case. This contrast demonstrates that higher-curvature corrections to the CV proposal are universally necessary, and they become significantly more pronounced at small $r_h/L$.

\subsection{General Time Dependence}
\label{sec:General_Time_Dependence}
Having established the late-time limits, we now numerically investigate the full time evolution of the complexity growth rate. Let $v_\infty$ and $v_{\mathrm{min}}$ denote the ingoing Eddington coordinate at the asymptotic boundary and the turning point, respectively. By symmetry, the turning point lies on the $t=0$ slice. Using eq.~\reef{dot_v} yields
\begin{equation}
\begin{aligned}
  \frac{\tau}{2} + r_{*}(r_{\infty}) - r_{*}(r_{\mathrm{min}}) 
  &= \int_{v_{\mathrm{min}}}^{v_{\infty}} \mathrm{d}v = \int_{r_{\mathrm{min}}}^{r_{\infty}} \frac{\dot{v}}{\dot{r}} \mathrm{d}r \\
  &= \int_{r_{\mathrm{min}}}^{r_\infty} \mathrm{d}r \left[ \frac{1}{f(r)} - \frac{P}{f(r) \dot{r} \left(1 - 2\tilde{\alpha} r^{-2d} \dot{r}^2 + \frac{2\tilde{\alpha} k}{r^2}\right)} \right] \,. 
\end{aligned}
\end{equation}
The boundary time $\tau$ can therefore be isolated as
\begin{equation}
\begin{aligned}
  \tau &= -2 \int_{r_{\mathrm{min}}}^{\infty} \mathrm{d}r \frac{P}{f(r) \dot{r} \left(1 - 2\tilde{\alpha} r^{-2d} \dot{r}^2 + \frac{2\tilde{\alpha} k}{r^2}\right)} \\
     &= -2 \int_{r_{\mathrm{min}}}^{\infty} \mathrm{d}r \frac{P}{f(r) \sqrt{-\mathcal{U}(P,r)} \left(1 + 2\tilde{\alpha} r^{-2d} \mathcal{U}(P,r) + \frac{2\tilde{\alpha} k}{r^2}\right)} \,,
\end{aligned}
\label{boundary_time}
\end{equation}
where $\mathcal{U}(P, r) = -\dot{r}^2$ as defined previously. We now analyze the divergence structure of the integral by rewriting the differential as
\begin{equation}
  \mathrm{d}\tau = \frac{-2PG(r)}{f(r) \sqrt{-\mathcal{U}(P,r)}} \, \mathrm{d}r \,, 
  \label{eq:isolated_form}
\end{equation}
where we have defined the factor
\begin{equation}
  G(r) \equiv \frac{1}{1 + 2\tilde{\alpha} r^{-2d}\mathcal{U}(P,r) + \frac{2\tilde{\alpha} k}{r^2}} \,.
\end{equation}
Since the apparent singularity at $r_{\mathrm{min}}^2 = -2\tilde{\alpha} k$ is safely avoided, $G(r)$ approaches a finite limit near the turning point $r_{\mathrm{min}}$ (where $\mathcal{U} \to 0$). Consequently, this factor does not alter the convergence properties of the integral. For turning points inside the horizon ($r_{\mathrm{min}} < r_h$), the integration path crosses $r_h$. Although $f(r) \simeq f'(r_h)(r-r_h)$ introduces an apparent logarithmic singularity, evaluating the integral via its Cauchy principal value perfectly cancels the contributions from both sides of the horizon, ensuring a finite result.

The true divergence of $\tau$, which signifies the late-time limit, arises solely from the $1/\sqrt{-\mathcal{U}(P,r)}$ factor at the lower integration limit. Expanding the effective potential around the turning point $r_{\mathrm{min}}$ (where $\mathcal{U} = 0$), we have
\begin{equation}
  \mathcal{U}(P,r) = \mathcal{U}'(r_{\mathrm{min}})(r-r_{\mathrm{min}}) + \frac{1}{2}\mathcal{U}''(r_{\mathrm{min}})(r-r_{\mathrm{min}})^2 + \dots \,.
\end{equation}
Generally, $\mathcal{U}'(r_{\mathrm{min}}) \neq 0$, meaning eq.~\eqref{eq:isolated_form} yields a finite $\tau$. However, at late times, $r_{\mathrm{min}}$ approaches a critical slice $\tilde{r}_{\mathrm{min}}$ where $\partial_r\mathcal{U}(P_{\infty}, \tilde{r}_{\mathrm{min}}) = 0$. In this limit, the potential vanishes to second order, $\mathcal{U}(P,r) \approx \mathcal{U}''(\tilde{r}_{\mathrm{min}})(r-\tilde{r}_{\mathrm{min}})^2 / 2$, leading to a logarithmic divergence in eq.~\eqref{eq:isolated_form}:
\beq
\tau \sim \int_{r \approx \tilde{r}_{\mathrm{min}}} \frac{\mathrm{d}r}{\sqrt{(r-\tilde{r}_{\mathrm{min}})^2}} \sim \int \frac{\mathrm{d}r}{r-\tilde{r}_{\mathrm{min}}} \sim \ln(r-\tilde{r}_{\mathrm{min}}) \to \infty \,. 
\eeq

\subsubsection*{Numerical Results}

\begin{figure}[htbp]
	\centering
	\includegraphics[scale=0.7]{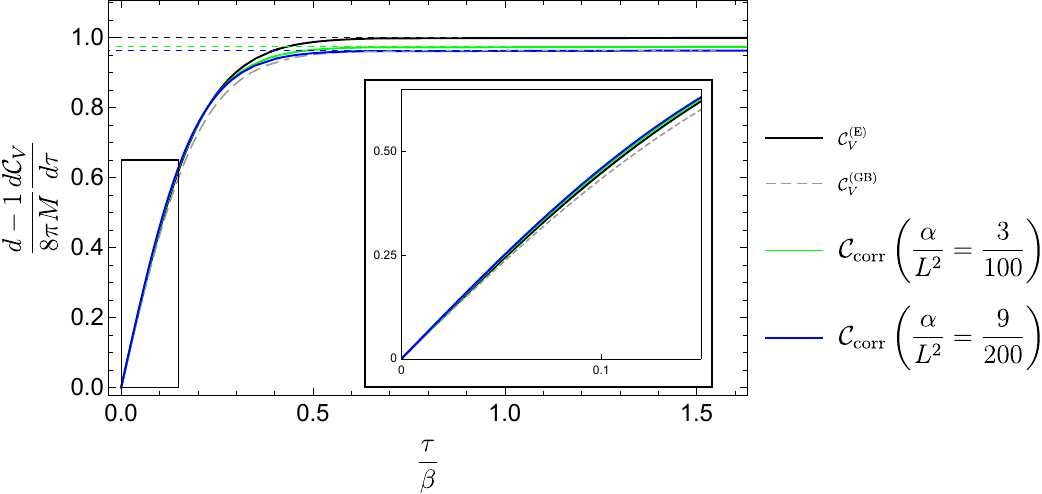}
	\caption{Time evolution of the complexity growth rate for the planar ($k=0$) Gauss-Bonnet black hole. The solid black, gray dashed, green, and blue curves correspond to pure Einstein gravity ($\alpha/L^2=0$), the uncorrected GB volume ($\alpha/L^2=9/200$), and the corrected GB volumes with $\alpha/L^2=3/100$ and $9/200$, respectively. As $\tau \to \infty$, the fully corrected result (blue) converges to the uncorrected geometric volume (gray) for a given coupling.}
\label{fig_Pv_tau_Vcorrected_planar_a}
\end{figure}

To complete our analysis, we numerically evaluate the full time dependence for various topologies ($k$) in $d=4$. Figure~\ref{fig_Pv_tau_Vcorrected_planar_a} displays the evolution for the planar topology ($k=0$). At early times, the fully corrected growth rates are largely insensitive to the coupling $\alpha/L^2$. During the intermediate time, however, they deviate significantly from the uncorrected CV proposal (pure geometric volume) and remain universally suppressed relative to Einstein gravity. This confirms the necessity of distinct corrections to the CV proposal even for planar black holes. Therefore, explicit comparisons with the uncorrected CV proposal are omitted hereafter.

\begin{figure}[tbp]
  \centering
  \begin{subfigure}[b]{0.48\textwidth}
    \centering
    \includegraphics[width=\textwidth]{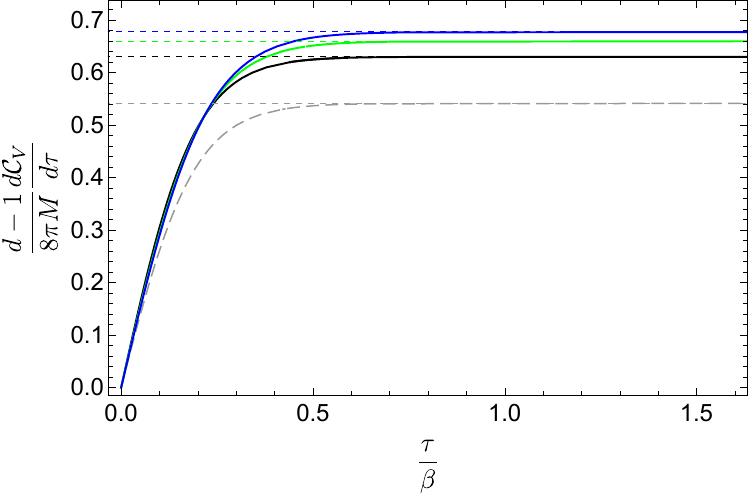}
    \caption{Time evolution ($r_h/L = 1$)}
    \label{fig:k1_rh1_global}
  \end{subfigure}
  \hfill
  \begin{subfigure}[b]{0.48\textwidth}
    \centering
    \includegraphics[width=\textwidth]{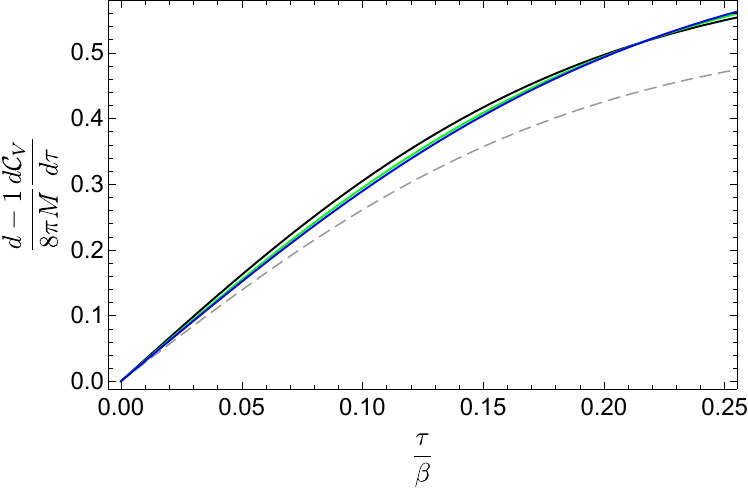}
    \caption{Early time competition ($r_h/L = 1$)}
    \label{fig:k1_rh1_early}
  \end{subfigure}
  
  \vspace{0.6cm}
  
  \begin{subfigure}[b]{0.48\textwidth}
    \centering
    \includegraphics[width=\textwidth]{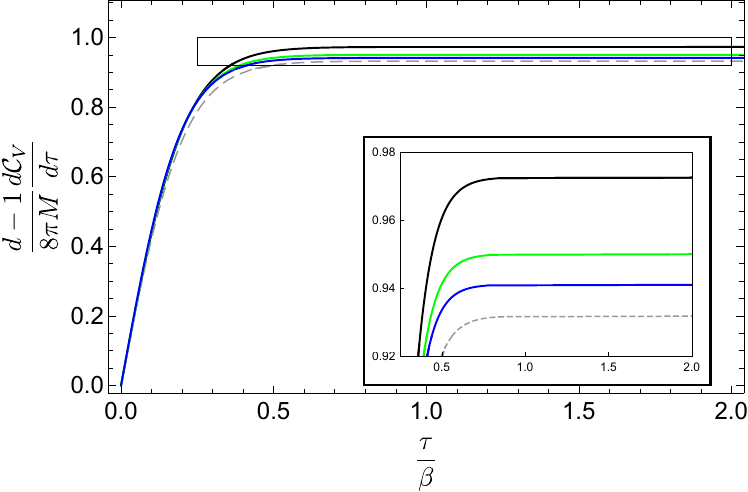} 
    \caption{Time evolution ($r_h/L = 5$)}
    \label{fig:k1_rh5_global}
  \end{subfigure}
  \hfill
  \begin{subfigure}[b]{0.48\textwidth}
    \centering
    \includegraphics[width=\textwidth]{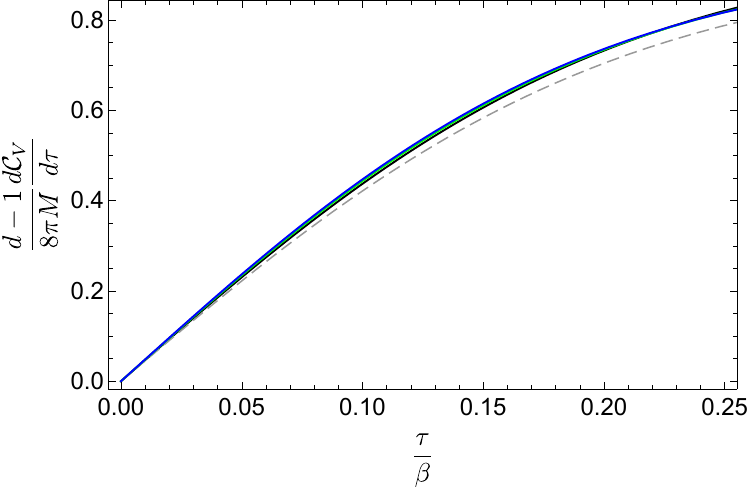} 
    \caption{Early time competition ($r_h/L = 5$)}
    \label{fig:k1_rh5_early}
  \end{subfigure}
  \caption{Evolution of the complexity growth rate for the spherical topology ($k=1$). The top and bottom rows correspond to small ($r_h/L=1$) and large ($r_h/L=5$) horizons, respectively. The left panels illustrate the global time evolution, while the right panels zoom in on the early time dynamics to highlight the competition effect. The solid black, gray dashed, green, and blue curves correspond to pure Einstein gravity ($\alpha/L^2=0$), the uncorrected GB volume ($\alpha/L^2=9/200$), and the corrected GB volumes with $\alpha/L^2=3/100$ and $9/200$, respectively.}
  \label{fig_Pv_tau_Vcorrected_spherical_a}
\end{figure}

\begin{figure}[tbp]
  \centering
  \begin{subfigure}[b]{0.48\textwidth}
    \centering
    \includegraphics[width=\textwidth]{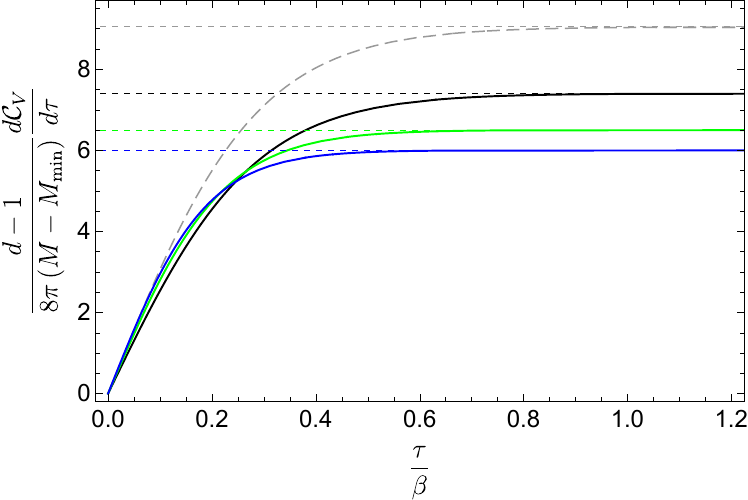}
    \caption{Time evolution ($r_h/L = 0.8$)}
    \label{fig:km1_rh1_global}
  \end{subfigure}
  
  \vspace{0.6cm}
  
  \begin{subfigure}[b]{0.48\textwidth}
    \centering
    \includegraphics[width=\textwidth]{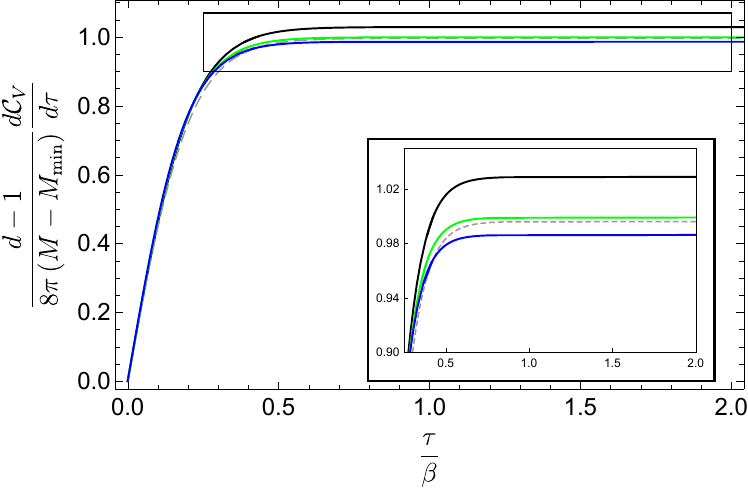} 
    \caption{Time evolution ($r_h/L = 5$)}
    \label{fig:km1_rh5_global}
  \end{subfigure}
  \hfill
  \begin{subfigure}[b]{0.48\textwidth}
    \centering
    \includegraphics[width=\textwidth]{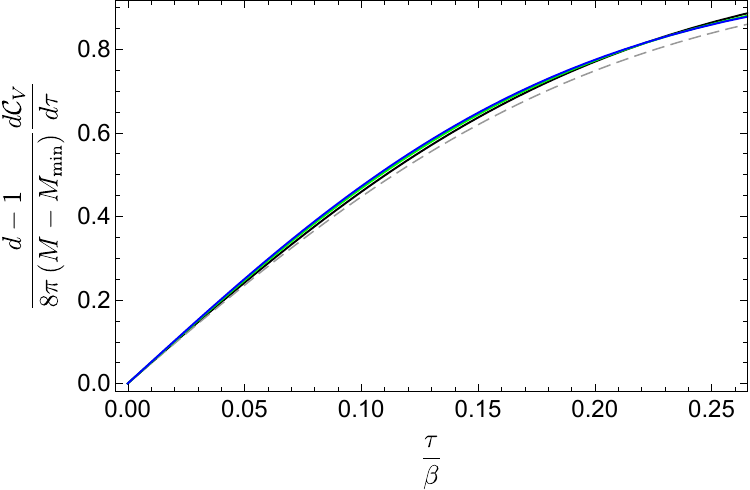} 
    \caption{Early time competition ($r_h/L = 5$)}
    \label{fig:km1_rh5_early}
  \end{subfigure}
  \caption{Evolution of the complexity growth rate for the hyperbolic topology ($k=-1$). The top panel shows the time evolution in the small horizon regime ($r_h/L=0.8$). The bottom panels display the large horizon regime ($r_h/L=5$), with the left panel illustrating the global evolution and the right panel zooming in on the early time dynamics to highlight the competition effect. The solid black, gray dashed, green, and blue curves correspond to pure Einstein gravity ($\alpha/L^2=0$), the uncorrected GB volume ($\alpha/L^2=9/200$), and the corrected GB volumes with $\alpha/L^2=3/100$ and $9/200$, respectively.}
  \label{fig_Pv_tau_Vcorrected_hyperbolic_a}
\end{figure}

The full time dependence for curved horizons is shown in figures~\ref{fig_Pv_tau_Vcorrected_spherical_a} and~\ref{fig_Pv_tau_Vcorrected_hyperbolic_a}. For spherical black holes ($k = 1$), a positive coupling enhances the late-time rate at small $r_h/L$, but suppresses it at large $r_h/L$. Conversely, for hyperbolic topologies ($k = -1$), the coupling universally suppresses the late-time limit across all $r_h/L$. Interestingly, for small hyperbolic black holes, the early time growth transiently exceeds the Einstein gravity baseline before crossing over and settling below it. In fact, this competition effect is quite universal for arbitrary topology, and such behavior marks a novel qualitative feature absent in the uncorrected CV proposal. Mathematically, this crossover reflects a dynamic sign reversal in the $\alpha$-derivative of the growth rate. The Gauss-Bonnet correction plays a nontrivial time-dependent role: accelerating early-time growth while suppressing late-time growth. Crucially, across all topologies, the results here perfectly align with the late-time behaviors obtained in section~\ref{sec:Late_Time_Behavior}, confirming both our analytical predictions and prior numerical limits.

\section{Holographic Complexity in Gauss-Bonnet AdS-Vaidya Spacetime}
\label{sec:Vaidya}
Building upon the static analysis, we now turn to dynamical shock wave geometries. The analytic description of a collapsing shell (the Vaidya geometry) was rigorously generalized to Gauss-Bonnet gravity by Kobayashi~\cite{Kobayashi:2005ch}. He demonstrated that replacing the static mass parameter with a time-dependent mass satisfies the field equations sourced by a null shell. Following the framework and notation of refs.~\cite{Chapman:2018dem,Chapman:2018lsv}, the line element of the $(d+1)$-dimensional Gauss-Bonnet AdS-Vaidya spacetime in ingoing Eddington-Finkelstein coordinates reads
\begin{equation}
  \mathrm{d}s^2 = - F(r,v) \mathrm{d}v^2 + 2 \mathrm{d}r \mathrm{d}v + r^2 \mathrm{d}\Sigma_{k,d-1}^2 \,, 
\label{shock wave_geometry}
\end{equation}
where the time-dependent emblackening factor $F(r,v)$ takes the form:
\begin{equation}
  F(r,v) = k + \frac{r^2}{2\tilde{\alpha}} \left( 1 - \sqrt{1 + 4\tilde{\alpha} \left( \frac{f_{\mathrm{p}}(v)}{r^d} - \frac{1}{L^2} \right)} \right) \,. 
  \label{eq:F_master}
\end{equation}
Compared to the static metric in eq.~\reef{eq:Eddington_Finkelstein_coordinates}, the static mass parameter $\omega^{d-2}$ is promoted to a dynamical mass profile $f_{\mathrm{p}}(v)$, which in the thin-shell limit along the null hypersurface $v=v_s$ takes the form:
\beq
f_{\mathrm{p}}(v) = \omega_1^{d-2} \big(1 - \mathcal{H}(v-v_s)\big) + \omega_2^{d-2} \mathcal{H}(v-v_s) \,, 
\label{heavyX}
\eeq
where $\mathcal{H}(v)$ denotes the Heaviside step function. This geometry matches an interior static black hole ($\omega_1$) to an exterior one ($\omega_2$) across the shock wave.
We will focus on planar and spherical black holes although the framework naturally extends to the hyperbolic case. The static thermodynamic relations in eqs.~\reef{M_omega_relation} and~\reef{temperature_entropy} remain valid within the piecewise static regions on either side of the shell.

The bulk Lagrangian for the complete volume in the full region reads:
\begin{equation}
\mathbf{L}_{\mathrm{full}} = \left( 1 + \frac{2\tilde{\alpha}}{(d-1)(d-2)} R_{\mathcal{B}} \right) \sqrt{ \dot{v} \bigl( 2\dot{r} - F(r, v) \dot{v} \bigr) } \, r^{d-1} \,,
  \label{eq:Lagrangian}
\end{equation}
where the Ricci scalar $R_{\mathcal{B}}$ evaluates to
\begin{equation}
  R_{\mathcal{B}} = \frac{d-1}{r^2 \mathcal{A}^2} \left[ \mathcal{A} \left( (d-2)(-\dot{r}^2 + k \mathcal{A}) - 2r\ddot{r} \right) + r\dot{r} \dot{\mathcal{A}} \right] \,.
\end{equation}
Here, $\mathcal{A} \equiv 2 \dot{r} \dot{v} - F(r,v) \dot{v}^2$. Accounting for the GHY boundary term, varying the full action yields the equations of motion:
\begin{align}
  \ddot{v} &= \frac{1}{\mathcal{D}_v} \bigg\{ 
  2(2d-3)\tilde{\alpha} r^{6d} - 2(d-2)\tilde{\alpha}(4k-F) r^{2+4d}\dot{v}^2 - 4(d-1) r^{4+4d}\dot{v}^2 \notag \\
  &\quad - 2\tilde{\alpha}F(4k-F) r^{4+2d}\dot{v}^4 + 2d\tilde{\alpha} F^3 r^6 \dot{v}^6 + \tilde{\alpha} r^{3+4d}\dot{v}^2 \partial_r F + 2 r^{7+2d}\dot{v}^4 \partial_r F \notag \\
  &\quad + 2\tilde{\alpha} r^{5+2d}\dot{v}^4 \left[ 2\partial_v F + (2k-F)\partial_r F \right] - \tilde{\alpha} F r^7 \dot{v}^6 \left( 4\partial_v F + 3F \partial_r F \right)
  \bigg\} \,, \\[10pt]
  \ddot{r} &= \frac{1}{\mathcal{D}_r} \bigg\{
  d\tilde{\alpha} r^{8d} - 2\tilde{\alpha} \left[ 2k - (d-1)F \right] r^{2+6d}\dot{v}^2 - 2(d-2)\tilde{\alpha} F(4k-F) r^{4+4d}\dot{v}^4 \notag \\
  &\quad - 4(d-1)F r^{6+4d}\dot{v}^4 - 2\tilde{\alpha} F^2 \left[ 2k - (d-1)F \right] r^{6+2d}\dot{v}^6 + d\tilde{\alpha} F^4 r^8 \dot{v}^8 \notag \\
  &\quad - 2 r^{9+2d}\dot{v}^6 \partial_v F + \tilde{\alpha} F^2 r^9 \dot{v}^8 \partial_v F + \tilde{\alpha} r^{3+6d}\dot{v}^2 \partial_r F - 2 r^{7+4d}\dot{v}^4 \partial_r F \notag \\
  &\quad - \tilde{\alpha} r^{7+2d}\dot{v}^6 \left[ (4k-2F)\partial_v F - F^2 \partial_r F \right] \notag \\
  &\quad + \tilde{\alpha} r^{5+4d}\dot{v}^4 \left[ \partial_v F + 2(-2k+F)\partial_r F \right]
  \bigg\} \,,
  \label{complete_equations_of_motion}
\end{align}
where the denominators are given by
\begin{align}
  \mathcal{D}_v &= 2 r^3 \left[ 3\tilde{\alpha} r^{4d} - 2 r^{2+2d}\dot{v}^2 \left( \tilde{\alpha}(2k+F) + r^2 \right) + 3\tilde{\alpha} F^2 r^4 \dot{v}^4 \right] \,, \\
  \mathcal{D}_r &= r^2 \dot{v}^2 \mathcal{D}_v \,.
\end{align}

Since we are interested in the behavior of $\dot{r}$ and $\dot{v}$ on both sides of the shock wave, we integrate the full equations of motion across a neighbourhood of the shock wave and subsequently take the width of this neighbourhood to zero. In this limit, only terms containing $\partial_v F$ survive:
\begin{align}
  \ddot{v} &\supset \frac{2\tilde{\alpha} \dot{v}^4 r^{5+2d}}{\mathcal{D}_v} \left[ 2 - 2 F r^{2-2d} \dot{v}^2 \right] \partial_v F \,, \label{singular_term}\\[10pt]
  \ddot{r} &\supset \frac{\tilde{\alpha} \dot{v}^4 r^{5+4d}}{\mathcal{D}_r} \bigg\{ 1 - \left[ \frac{4k-2F}{r^{2d}} + \frac{2}{\tilde{\alpha} r^{4d-2}} \right] r^2 \dot{v}^2 + \frac{F^2}{r^{4d-4}} \dot{v}^4 \bigg\} \partial_v F \,,
\end{align}
where the notation $\supset$ indicates that we retain only the singular terms proportional to $\partial_v F$. Note that these equations of motion exactly reproduce the standard results in the Einstein limit ($\tilde{\alpha} \to 0$). However, here both the jump in $\dot{r}$ and $\dot{v}$ appear, which is quite distinct from the Einstein case.

Proceeding with a perturbative expansion in the coupling $\tilde{\alpha}$, let $f^{(0)}$ and $\dot{v}^{(0)}$ denote the emblackening factor and the continuous $\dot{v}$ in the Einstein limit, respectively. In the remainder of this paper, we work strictly to linear order in the Gauss-Bonnet coupling. For notational simplicity, equality signs ($=$) in perturbative expressions are understood to hold up to $\mathcal{O}(\tilde{\alpha})$, and higher-order terms are implicitly dropped. Changing the integration variable in eq.~\reef{singular_term} from $\lambda$ to $F$, we integrate across the shock wave to find the leading-order $\mathcal{O}(\tilde{\alpha})$ discontinuity
\begin{equation}
  \Delta \dot{v} =\dot{v}_2-\dot{v}_1= \frac{\tilde{\alpha} \dot{v}^{(0)}(r_s) \Delta f^{(0)}}{2 r_s^{2(d+1)}} \left[ -2 r_s^{2d} + r_s^2 (\dot{v}^{(0)}(r_s))^2 \Sigma f^{(0)} \right] \,, \label{eq:Delta_v_explicit}
\end{equation}
where $r_s$ marks the intersection radius of the extremal surface with the collapsing shell. For brevity, we have defined $\Delta f^{(0)} \equiv f_{2}^{(0)}(r_s) - f_{1}^{(0)}(r_s)$ and $\Sigma f^{(0)} \equiv f_{1}^{(0)}(r_s) + f_{2}^{(0)}(r_s)$. Eliminating $\dot{r}$ from eq.~\reef{eq:Conserved_momentum} using the gauge constraint~\reef{eq:gauge_constraint}, we expand the conserved momentum $P_i$ to first order in $\tilde{\alpha}$:
\begin{equation}
\begin{split}
  P_i &= \frac{r_s^{2(d-1)} - f_i(r_s) \dot{v}_i(r_s)^2}{2 \dot{v}_i(r_s)} - \frac{\tilde{\alpha} \left[ r_s^{2d} - r_s^2 f_i^{(0)}(r_s) (\dot{v}_i^{(0)}(r_s))^2 \right]}{4 r_s^{2(d+3)} (\dot{v}_i^{(0)}(r_s))^3} \\
  &\quad \times \left[ r_s^{4d} - 2 r_s^{2(d+1)} \big( 2k - f_i^{(0)}(r_s) \big) (\dot{v}_i^{(0)}(r_s))^2 + r_s^4 (f_i^{(0)}(r_s))^2 (\dot{v}_i^{(0)}(r_s))^4 \right] \,.
\end{split}
\label{eq:P_i_general}
\end{equation}
Structurally, the first term in this expression retains the standard Einstein gravity form, albeit evaluated with the full GB metric and with $\dot{v}$. This term requires a more detailed analysis to examine its concrete form when expanded to order $\alpha$. Conversely, the second term isolates the explicit higher-derivative corrections, where all functions are strictly evaluated at zeroth order.

To proceed, we define the tortoise coordinate $r^*$ within the piecewise static patches via the relation $\mathrm{d}r^* = \mathrm{d}r/f(r)$. Imposing the boundary condition that $r^*$ vanishes at the AdS boundary, we obtain
\begin{equation}r^*(r) = -\int^\infty_r \frac{\mathrm{d}\tilde{r}}{f(\tilde{r})} . 
\label{tortoise_coordinates}\end{equation}
We define the outgoing null coordinate $u$ as $u \equiv v - 2 r^*(r)$. It is important to note that while the coordinates $v$ and $r$ are globally continuous, the mass jump across the shell induces a discontinuity in $r^*$, and consequently in $u$ and $t$. By convention, we draw a `Penrose-like' diagram in figure~\ref{fig_eternal_shock_BH}, which is drawn such that $v$, $u$, and $t$ appear continuous at the expense of a discontinuous $r$ coordinate.

In the following, we apply the corrected volume functional to two distinct scenarios: the formation of a one-sided black hole in section~\ref{sec:One-Sided_Black_Hole} and the shock wave perturbation of an eternal black hole in section~\ref{sec:Two-Sided_Black_Hole}.

\subsection{One-Sided Black Hole}
\label{sec:One-Sided_Black_Hole}
In this section, we consider a null shell collapsing into the pure AdS vacuum, resulting in the formation of a one-sided black hole. From the boundary CFT perspective, this is dual to a quantum quench~\cite{Das:2010yw,Abajo-Arrastia:2010ajo,Balasubramanian:2010ce,Balasubramanian:2011ur,Das:2014jna,Das:2014hqa,Das:2016lla}. The system is initially in the vacuum state. At $t=0$, a homogeneous energy injection creates an excited state.
\begin{figure}[htbp]
	%\vspace{2ex}
	\centering
	\includegraphics[scale=0.7]{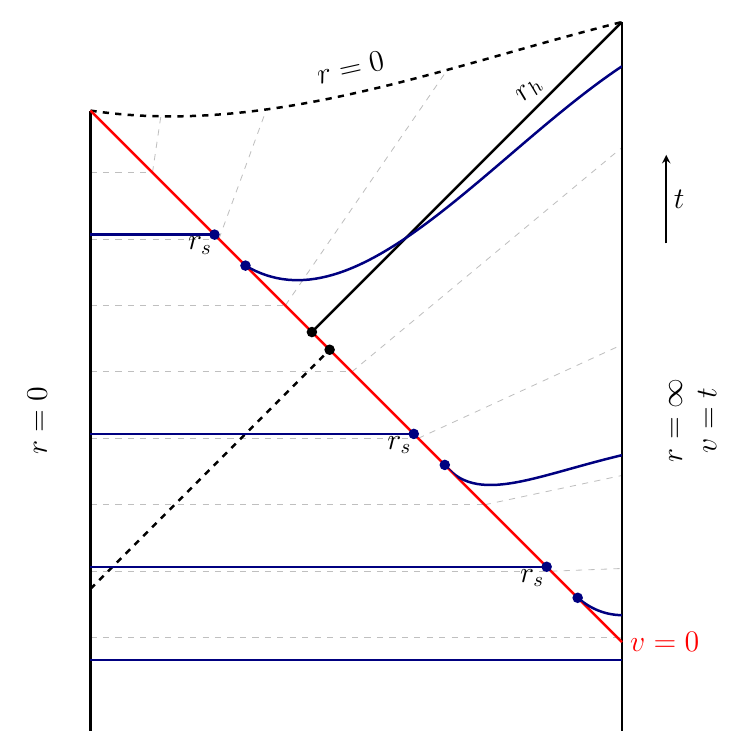}
	\caption{Penrose diagram illustrating the formation of a one-sided Gauss-Bonnet AdS black hole via a collapsing null shell (red line, $v=0$). Thin dashed gray lines denote constant time slices. To avoid a conical singularity at $r=0$, the maximal volume surfaces (blue curves) coincide with these slices within the vacuum region. The solid and thick-dashed black lines represent the event horizon and its extension into the vacuum, respectively. Adapted from~\cite{Chapman:2018dem}.}
\label{fig__one-sided_BH}
\end{figure} 
Figure~\ref{fig__one-sided_BH} illustrates the maximal volume surfaces. Here $f_{\mathrm{p}}(v)$ given in eq.~\eqref{heavyX} simplifies to $f_{\mathrm{p}}(v) = \omega^{d-2} \, \mathcal{H}(v)$. The spacetime is separated into two distinct static regions by the shock wave at $v=0$:
\begin{eqnarray}
v<0 \ :&& \qquad F(r, v) = f_{\mathrm{vac}}(r) = k + \frac{r^2}{2\tilde{\alpha}} \left( 1 - \sqrt{1 - \frac{4\tilde{\alpha}}{L^2}} \right) \,, \label{eq:fVac}\\
v>0\ :&& \qquad F(r, v) = f_{\mathrm{BH}}(r) = k + \frac{r^2}{2\tilde{\alpha}} \left( 1 - \sqrt{1 + 4\tilde{\alpha} \left( \frac{\omega^{d-2}}{r^d} - \frac{1}{L^2} \right)} \right) \,. \label{eq:fBH}
\end{eqnarray}

Two boundary conditions are required to determine this spacelike hypersurface. The first boundary condition is imposed to avoid a conical singularity at $r=0$, which yields $\dot{t} = \dot{v} - \dot{r}/f = 0$. Consequently, eq.~\reef{eq:Conserved_momentum} dictates that the conserved momentum in the vacuum region vanishes ($P_{\mathrm{vac}} = 0$). 
We expand the physical branch of $\dot{r}^2$ discussed previously (eq.~\eqref{eq:Y2}) to order $\mathcal{O}(\tilde{\alpha})$:
\begin{equation}
\dot{r}^2 = P^2 + r^{2(d-1)} f_i(r) + \frac{4 \tilde{\alpha} P^2}{r^2} \left[ f_i^{(0)}(r) - k + P^2 r^{-2(d-1)} \right] + \mathcal{O}(\tilde{\alpha}^2) \,. 
\label{eq:r_dot_squared_expansion}
\end{equation}
Evaluating this expression in the vacuum region ($P_{\mathrm{vac}} = 0$) yields
\begin{equation}\label{eq:dotr_vac}
\dot{r}_{\mathrm{vac}}(r_s) = r_s^{d-1} \sqrt{ f_{\mathrm{vac}}(r_s)} \,. 
\end{equation}
Moreover, setting $P_{\mathrm{vac}}=0$ in eq.~\reef{dot_v} eliminates the higher-derivative corrections, reducing the relation $\dot{v}_{\mathrm{vac}} = \dot{r}_{\mathrm{vac}}/f_{\mathrm{vac}}$. This yields
\begin{equation}\label{dotv_vac}
 \qquad \dot{v}_{\mathrm{vac}}(r_s) = \frac{\dot{r}(r_s)}{f_{\mathrm{vac}}(r_s)}=\frac{r_s^{d-1}}{\sqrt{f_{\mathrm{vac}}(r_s)}} \,.
\end{equation}
Substituting the zeroth-order profile $\dot{v}^{(0)}(r_s) = r_s^{d-1}/\sqrt{f_{\mathrm{vac}}^{(0)}(r_s)}$ into the jump for $\dot{v}$ in eq.~\reef{eq:Delta_v_explicit}, $\dot{v}$ on the black hole side of the shell evaluates to
\begin{equation}\dot{v}_{\mathrm{BH}}(r_s) = \frac{r_s^{d-1}}{\sqrt{f_{\mathrm{vac}}(r_s)}} + \frac{\tilde{\alpha} r_s^{d-3} \left(\Delta f^{(0)}(r_s)\right)^2}{2 \left( f_{\mathrm{vac}}^{(0)}(r_s) \right)^{3/2}} \,.\label{eq:v_dot_BH_final}\end{equation} 
Substituting eq.~\reef{eq:v_dot_BH_final} into eq.~\reef{eq:P_i_general}, the conserved momentum in the black hole region expands to $\mathcal{O}(\tilde{\alpha})$ as
\begin{equation}
\begin{split}
  P_{\mathrm{BH}} &=- \frac{r_s^{d-1} \Delta f}{2 \sqrt{f_{\mathrm{vac}}}} - \frac{\tilde{\alpha} r_s^{d-3} \Sigma f^{(0)} (\Delta f^{(0)})^2}{4 (f_{\mathrm{vac}}^{(0)})^{3/2}} - \frac{\tilde{\alpha} \left[ r_s^{2d} - r_s^2 f_{\mathrm{BH}}^{(0)} (\dot{v}^{(0)})^2 \right]}{4 r_s^{2(d+3)} (\dot{v}^{(0)})^3} \\
  &\quad \times \left[ r_s^{4d} - 2 r_s^{2(d+1)} (2k - f_{\mathrm{BH}}^{(0)}) (\dot{v}^{(0)})^2 + r_s^4 (f_{\mathrm{BH}}^{(0)})^2 (\dot{v}^{(0)})^4 \right] \,,
\end{split}
\label{eq:P_BH_Full_Expansion}
\end{equation}
where $\Delta f \equiv f_{\mathrm{BH}} - f_{\mathrm{vac}}$ and $\Sigma f^{(0)} \equiv f_{\mathrm{BH}}^{(0)} + f_{\mathrm{vac}}^{(0)}$. The leading term incorporates the full Gauss-Bonnet metric. For brevity, its explicit $\mathcal{O}(\tilde{\alpha})$ expansion is omitted here.

\begin{figure}[htbp]
\centering
\includegraphics[scale=0.8]{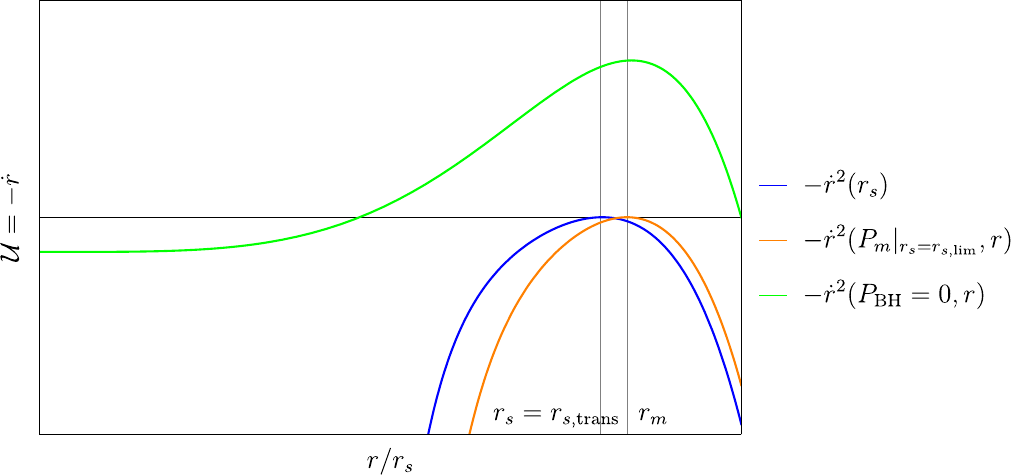}
\caption{The effective potential $\mathcal{U} = -\dot{r}^2$ as a function of $r/r_s$ for a one-sided black hole in $d=4$ Gauss-Bonnet gravity. The green curve corresponds to the potential for the initial hypersurface with zero momentum ($P_{\mathrm{BH}}=0$). The blue curve represents the potential whose peak is exactly tangent to the $\mathcal{U}=0$ line at $r_s=r_{s,\mathrm{trans}}$, marking the transition point where the direction of $\dot{r}_{\mathrm{BH}}(r_s)$ flips. The orange curve illustrates the critical potential evaluated at the limiting shock wave position $r_{s,\mathrm{lim}}$ with momentum $P_m$. This curve is tangent to the zero axis at the bottleneck $r=r_m$, corresponding to the late-time limit where the boundary time diverges.} 
\label{fig:Potential}
\end{figure}

The second boundary condition anchors the surface to the asymptotic boundary at time $t_0 > 0$:
\begin{equation}
  t_0 = \int_{r_s}^{\infty} \frac{\dot{v}}{\dot{r}} \mathrm{d}r = \int_{r_s}^{\infty} \frac{\mathrm{d}r}{f_{\mathrm{BH}}(r)} \left( 1 - \frac{P_{\mathrm{BH}}}{\sqrt{-\mathcal{U}(P_{\mathrm{BH}},r)} \left(1 + 2\tilde{\alpha} r^{-2d} \mathcal{U}(P_{\mathrm{BH}},r) + \frac{2\tilde{\alpha} k}{r^2}\right)} \right) \,. 
\label{The_second_boundary_condition}
\end{equation}
To successfully reach the asymptotic boundary, the surface must satisfy the condition $\mathcal{U} \equiv -\dot{r}_{\mathrm{BH}}^2 \le 0$ everywhere. We define the critical momentum $P_m$ and the location of the potential maximum $r_m$ via the conditions:
\begin{equation}
  \partial_r \mathcal{U}(P_m, r_m) = 0 \,, \qquad \mathcal{U}(P_m, r_m) = 0 \,.
\label{maximal_surf}
\end{equation}
Figure~\ref{fig:Potential} illustrates this effective potential $\mathcal{U}$ as a function of $r/r_s$ in $d=4$ Gauss-Bonnet gravity. For insufficiently large momentum $P_{\mathrm{BH}}$, the potential barrier is strictly positive ($\mathcal{U} > 0$), preventing the trajectory from escaping to the boundary and forcing it into the singularity at $r=0$. A critical geometric transition occurs at $r_s = r_{s,\mathrm{trans}}$, where the potential peak (blue curve) is exactly tangent to the $\mathcal{U}=0$ axis. At this specific shock wave position, $\dot{r}_{\mathrm{BH}}(r_s)$ vanishes and reverses direction. To ensure the extremal surfaces reach the asymptotic boundary, we require $P_{\mathrm{BH}}^2 \ge P_m^2$. Saturating this inequality (orange curve) corresponds to the effective potential becoming tangent to the $\mathcal{U}=0$ axis at $r=r_m$. Approaching this critical condition ($P_{\mathrm{BH}}^2 \to P_m^2$ and $r_s \to r_{s,\mathrm{lim}}$) yields a logarithmic divergence in the boundary time $t_0$. Physically, this divergence captures the late-time limit, where the holographic complexity exhibits its characteristic linear growth. 

Thus, the complete volume functional is given by
\begin{equation}
\begin{split}
W &= \Omega_{k,d-1} \left[ \int_0^{r_s} \frac{\mathbf{L}_{\mathrm{vac}}}{\dot{r}} \,\mathrm{d}r + \int_{r_s}^\infty \frac{\mathbf{L}_{\mathrm{BH}}}{\dot{r}}\,\mathrm{d}r \right] + W_{\mathrm{bdy}} \\
&= \Omega_{k,d-1} \left[ \int_0^{r_s} \frac{\mathbf{L}_{\mathrm{vac}} }{\dot{r}}\,\mathrm{d}r + \int_{r_s}^{r_{\mathrm{max}}} \left[ \frac{\mathbf{L}_{\mathrm{BH}} - P_{\mathrm{BH}}\dot{v} }{\dot{r}}\right] \mathrm{d}r + P_{\mathrm{BH}} t_0 \right] + W_{\mathrm{bdy}} \,.
\end{split}
\label{eq:W_total}
\end{equation}
Differentiating $W$ with respect to the boundary time $t_0$, and noting that both $r_s$ and $P_{\mathrm{BH}}$ depend on $t_0$, we obtain
\begin{equation}
  \frac{\mathrm{d}W}{\mathrm{d}t_0} = \frac{\partial W}{\partial t_0} + \frac{\partial W}{\partial P_{\mathrm{BH}}} \frac{\mathrm{d}P_{\mathrm{BH}}}{\mathrm{d}t_0} + \frac{\partial W}{\partial r_s} \frac{\mathrm{d}r_s}{\mathrm{d}t_0} \,.
  \label{chain_rule}
\end{equation}
The first two terms on the right-hand side of eq.~\reef{chain_rule} precisely mirror the eternal black hole analysis in section~\ref{sec:Eternal_Gauss-Bonnet_Black_Hole}, evaluating to $\Omega_{k,d-1} P_{\mathrm{BH}}$. The third term, arising from the shock wave junction, requires careful examination.

In the Ostrogradsky formalism, the boundary variations ($\Pi_v \delta \dot{v}$ and $\Pi_r \delta \dot{r}$) from the bulk are exactly canceled by the variation of $W_{\mathrm{bdy}}$. The total variation at the junction $r_s$ thus reduces to $\delta W|_{r_s} = [ P_r \delta r + P_v \delta v ]_{r_s^-}^{r_s^+}$, where $r_s^\pm$ denote the limits approaching the shell from the black hole ($+$) and vacuum ($-$) sides. Since the shock wave lies on a fixed null surface ($v = v_s$), the $v$-coordinate is rigidly constrained, yielding $\delta v|_{r_s} = 0$. Noting that $\delta r = \delta r_s$ at the junction, the variation simplifies to $\delta W|_{r_s} = ( P_r|_{\mathrm{BH}} - P_r|_{\mathrm{vac}} ) \delta r_s$.

For the variational principle to hold ($\delta W = 0$) under arbitrary displacements $\delta r_s$, this coefficient must vanish, yielding the Weierstrass-Erdmann corner condition: $[P_r] \equiv P_r|_{\mathrm{BH}} - P_r|_{\mathrm{vac}} = 0$. This continuity of the canonical radial momentum governs the refraction of the extremal surface across the shock wave.

In Einstein gravity, the gauge condition trivially reduces $P_r$ to $\dot{v}$, forcing $\dot{v}$ to be continuous. In Gauss-Bonnet gravity, however, the radial momentum reads
\begin{equation}
P_r = \left[ 1 + \frac{\tilde{\alpha}}{r^2} \left( 2k - f - \frac{r f'}{d-2} \right) \right] \dot{v} - \frac{\tilde{\alpha} f \dot{v}^3}{r^{2d}} \left( \frac{f}{2} + \frac{r f'}{d-2} \right) - \frac{\tilde{\alpha} r^{2d-4}}{2 \dot{v}} \,.
\label{eq:P_rGB} 
\end{equation}
Because $P_r$ depends explicitly on the metric function $f(r)$, which jumps across the shell, maintaining $[P_r] = 0$ necessitates a compensating discontinuity in $\dot{v}$. This jump manifests the refraction effect.

Consequently, the junction term $\partial W/\partial r_s$ strictly vanishes. This proves that the complexity growth rate is dictated solely by the conserved momentum in the black hole region, despite the higher-derivative $\tilde{\alpha}$ corrections and the discontinuities in $\dot{v}$ and $\dot{r}$:
\begin{equation}
\frac{\mathrm{d}\cv}{\mathrm{d}t_0} = \frac{1}{G_{\mathrm{bulk}} L}\,\frac{\mathrm{d}W}{\mathrm{d}t_0} = \frac{\Omega_{k,d-1}}{G_{\mathrm{bulk}} L} \,P_{\mathrm{BH}}\,. 
\label{VolVadFinal}
\end{equation}
Notably, as in Einstein gravity, the complexity growth rate remains directly proportional to the conserved momentum in the black hole region~\cite{Chapman:2018dem}. Furthermore, this feature exhibits parallels to the behavior of eternal black holes~\cite{Carmi:2017jqz}.

Specifically, the planar geometry admits an exact analytical treatment for eqs.~\reef{The_second_boundary_condition} and~\reef{eq:P_BH_Full_Expansion}. In this geometry, the expressions simplify significantly, yielding explicit results for the conserved momentum $P_{\mathrm{BH}}$ and the complexity growth rate:
\begin{equation}
  P_{\mathrm{BH}} = \frac{r_h^d}{2 L} \left( 1 - \frac{\tilde{\alpha}}{2 L^2} \right) \,, \qquad \frac{\mathrm{d}\cv}{\mathrm{d}t_0} = \frac{8 \pi M}{d-1} \left( 1 - \frac{\tilde{\alpha}}{2 L^2} \right) \,. 
  \label{eq:planar_GB_results}
\end{equation}
Several key physical features emerge from these results. First, as in Einstein gravity, the complexity grows linearly at a constant rate for all $t_0 > 0$. Second, a positive Gauss-Bonnet coupling $\tilde{\alpha}$ suppresses this growth rate relative to the Einstein prediction, exactly mirroring the late-time limit of the eternal black hole in eq.~\reef{eq:expansion_geometric}.
\subsubsection*{Early Time Limit}
We first examine the early time limit ($t_0 \to 0$). This condition corresponds to $r_s \to \infty$. Evaluating the conserved momentum in this limit yields
\begin{equation}\lim_{t_0\rightarrow 0}P_{\mathrm{BH}} = \frac{L \, \omega^{d-2}}{2} \left( 1 - \frac{\tilde{\alpha}}{2 L^2} \right) \,,\label{kapop_GB}\end{equation}
which translates to an initial complexity growth rate of
\begin{equation}\lim_{t_0\rightarrow 0} \frac{\mathrm{d}\cv}{\mathrm{d}t_0} = \frac{8\pi M}{d-1} \left( 1 - \frac{\tilde{\alpha}}{2L^2} \right) \,.\label{kapop2_GB}\end{equation} 
As in Einstein gravity, the growth rate undergoes an instantaneous jump at $t_0=0$ to a finite, albeit $\tilde{\alpha}$-suppressed, value. Crucially, this initial rate is entirely independent of the curvature $k$, perfectly reproducing the planar result in eq.~\eqref{eq:planar_GB_results}. 
\subsubsection*{Late-Time Limit}

In the late-time limit ($t_0 \to \infty$), the conserved momentum saturates its critical bound ($P_{\mathrm{BH}} \to P_m$) and $r_s$ reaches its minimal value $r_{s,\mathrm{lim}}$. Geometrically, the extremal surface wraps around the critical slice $r=r_m$, yielding a complexity growth rate structurally identical to that in Einstein gravity:
\begin{equation}
  \lim_{t_0 \to \infty} \frac{d-1}{8\pi M} \frac{\mathrm{d}\cv}{\mathrm{d}t_0} = \frac{2 P_m}{\omega^{d-2} L} \,.
  \label{eq:cv:late_rate_GB}
\end{equation} 
Although closed-form solutions for $P_m$ are generally intractable, applying the high temperature (small $L/r_h$) expansion directly recovers the same result as the eternal case: the critical radius $r_m$ (eq.~\eqref{eq:rf_expansion}) and the late-time rate (eq.~\eqref{eq:expansion_of_late_time_limit2}). Comparing this late-time behavior with the early time limit~\eqref{kapop2_GB} highlights a stark topological contrast: while the planar growth rate ($k=0$) remains strictly constant at all times, the spherical rate ($k=1$) evolves towards saturation (see the left panel of figure~\ref{fig:ComplexityGrowth}), driven by the interplay between curvature $k$ and the higher-derivative $\tilde{\alpha}$ corrections.
\subsubsection*{Numerical Results}

\begin{figure}[htbp]
\centering
 \includegraphics[width=0.48\textwidth]{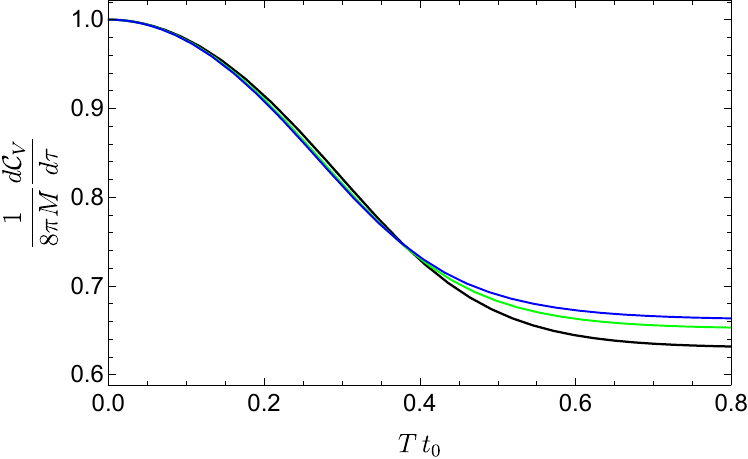}
 \hfill
 \includegraphics[width=0.48\textwidth]{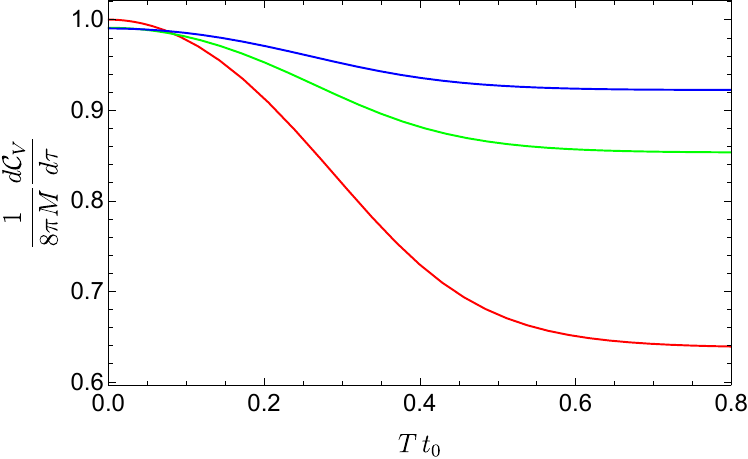}
\caption{Dimensionless holographic complexity growth rate versus boundary time $T t_0$ for a $d=4$ spherical ($k=1$) Gauss-Bonnet black hole. 
Left: Varying the Gauss-Bonnet coupling at fixed horizon radius ($r_h/L=1$). Curves represent Einstein gravity ($\alpha/L^2=0$, solid black) and Gauss-Bonnet gravity with $\alpha/L^2=3/100$ (dashed green) and $9/200$ (dot-dashed blue). 
Right: Varying the horizon radius at fixed Gauss-Bonnet coupling ($\alpha/L^2=1/100$). Curves represent $r_h/L=1$ (solid red), $2$ (dashed green), and $3$ (dot-dashed blue). In all configurations, the numerical curves smoothly asymptote to their respective analytical late-time saturation limits.}
\label{fig:ComplexityGrowth}
\end{figure}

We numerically evaluate the full time evolution of the holographic complexity growth rate for a $d=4$ spherical ($k=1$) black hole. Figure~\ref{fig:ComplexityGrowth} illustrates this dimensionless rate as a function of the dimensionless boundary time $T t_0$. Consistent with our analytical derivations and the behavior in Einstein gravity, the growth rate in all Gauss-Bonnet configurations initially decreases, eventually approaching its late-time saturation limit from above. 

The left panel of figure~\ref{fig:ComplexityGrowth} isolates the effect of the Gauss-Bonnet coupling at a fixed horizon radius ($r_h/L=1$). Interestingly, the curves nearly coincide at early times, implying that higher-derivative corrections are initially negligible. At late times, however, these effects become pronounced: a larger coupling $\alpha/L^2$ noticeably enhances the saturation value relative to the Einstein gravity baseline. The right panel shows the effect of varying the horizon size $r_h/L$ at a fixed coupling ($\alpha/L^2=1/100$). As shown, larger black holes attain significantly higher dimensionless growth rates at late times.

\subsection{Two-Sided Black Hole}
\label{sec:Two-Sided_Black_Hole}

\begin{figure}[htbp]
	\centering
	\includegraphics[scale=0.7]{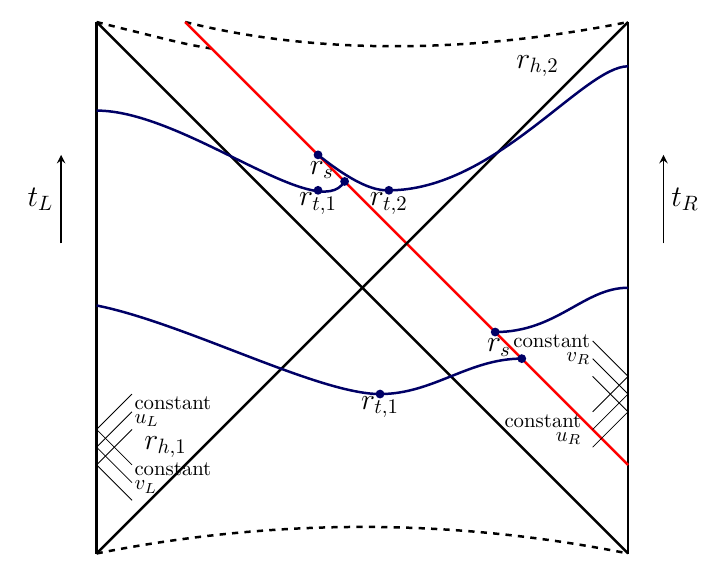}
	\caption{Penrose diagram of extremal surfaces in a two-sided Gauss-Bonnet AdS black hole perturbed by a shock wave. Key geometric features are marked, including the turning points ($r_{t,1}$ and $r_{t,2}$) and the shell-crossing point $r_s$. Lines of constant null coordinates ($u, v$) and boundary time flows are also indicated. Adapted from~\cite{Chapman:2018lsv}.
}

	\label{fig_eternal_shock_BH}
\end{figure}
Based on eq.~\reef{TFDState}, we now proceed to construct the holographic dual of a perturbed TFD state. Specifically, we investigate the time-dependent geometry of a null shock wave injected into an eternal black hole background with arbitrary energy (see figure~\ref{fig_eternal_shock_BH}). This Vaidya metric models thermal quenches~\cite{Buchel:2012gw,Buchel:2013lla}. In the boundary CFT language, this bulk perturbation translates to a perturbed state $| \Psi_{\mathrm{TFD}}\rangle_{\mathrm{pert}}$ created by acting on the TFD state with a specific precursor operator $\mathcal{O}_{\mathrm{R}}$ inserted at an early time $v_s=-t_w$ on the right boundary~\cite{Shenker:2013pqa,Shenker:2013yza}:
\begin{equation}
|\Psi_{\mathrm{TFD}}\rangle_{\mathrm{pert}}= \mathcal{O}_{\mathrm{R}}(-t_w)\, | \Psi_{\mathrm{TFD}} \rangle = U_{\mathrm{R}}(t_w)\, \mathcal{O}_{\mathrm{R}}\, U^{\dag}_{\mathrm{R}}(t_w)\, | \Psi_{\mathrm{TFD}} \rangle \,.
  \label{eq:perturbed_state}
\end{equation} 
Physically, this operation corresponds to evolving the right CFT backward by $t_w$, inserting the operator, and evolving forward again. At arbitrary boundary times $t_{\mathrm{L}}$ and $t_{\mathrm{R}}$, the state evolves as
\begin{equation}
\begin{split}
|\Psi_{\mathrm{TFD}}(t_{\mathrm{L}}, t_{\mathrm{R}})\rangle_{\mathrm{pert}} &= U_{\mathrm{L}}(t_{\mathrm{L}}) \, U_{\mathrm{R}}(t_{\mathrm{R}}) \, |\Psi_{\mathrm{TFD}}\rangle_{\mathrm{pert}} \\
  &= U_{\mathrm{L}}(t_{\mathrm{L}}) \, U_{\mathrm{R}}(t_{\mathrm{R}}+t_w)\, \mathcal{O}_{\mathrm{R}}\, U^{\dag}_{\mathrm{R}}(t_w)\, | \Psi_{\mathrm{TFD}} \rangle \\
  &= U_{\mathrm{R}}(t_{\mathrm{R}}+t_w)\, \mathcal{O}_{\mathrm{R}}\, U_{\mathrm{R}}(t_{\mathrm{L}}-t_w) \, | \Psi_{\mathrm{TFD}} \rangle \,, 
\end{split}
\label{eq:perturbed_time_evolution}
\end{equation} 
where we have utilized the commutativity of the left and right operators and the boost invariance of the TFD state. Evaluating the complexity of such precursor states typically reveals subtle cancellation patterns, a phenomenon known as the ``switchback effect''~\cite{Susskind:2014jwa,Stanford:2014jda}. While injecting the shock wave at a fixed time $-t_w$ breaks the original invariance of the eternal black hole, a generalized time-shift symmetry emerges. Specifically, the time-evolved state in eq.~\reef{eq:perturbed_time_evolution} remains invariant under the transformation:
\begin{equation}
  t_{\mathrm{L}} \to t_{\mathrm{L}} + \delta t \,, \quad t_{\mathrm{R}} \to t_{\mathrm{R}} - \delta t \,, \quad \text{and} \quad t_w \to t_w + \delta t \,.
\label{eq:generalized_symmetry}
\end{equation} 
Consequently, the holographic complexity only depends on $t_{\mathrm{R}} + t_w$ and $t_{\mathrm{L}} - t_w$.

The metric function $F(r,v)$ in eq.~\reef{shock wave_geometry} undergoes a step-function transition across the shock wave:
\begin{eqnarray}
v_{\mathrm{R}}< - t_{w} \ :&& \qquad F(r, v) = f_{1} (r) = k + \frac{r^2}{2\tilde{\alpha}} \left( 1 - \sqrt{1 + 4\tilde{\alpha} \left( \frac{\omega_1^{d-2}}{r^d} - \frac{1}{L^2} \right)} \right) \,,\label{fBH1}\\
v_{\mathrm{R}}> - t_{w}\ :&& \qquad F(r, v) = f_{2} (r) = k + \frac{r^2}{2\tilde{\alpha}} \left( 1 - \sqrt{1 + 4\tilde{\alpha} \left( \frac{\omega_2^{d-2}}{r^d} - \frac{1}{L^2} \right)} \right) \,,\label{eq:right_metric_profile}
\end{eqnarray} 
where $f_1(r)$ and $f_2(r)$ correspond to the blackening factors for masses $M_1$ and $M_2$, respectively. Following the framework of~\cite{Chapman:2018lsv}, we designate the geometries before and after the transition as $\mathrm{BH}_1$ and $\mathrm{BH}_2$.

We consider extremal surfaces anchored at boundary times $t_{\mathrm{L}}$ and $t_{\mathrm{R}}$, with the left and right boundaries satisfying $v_{\mathrm{L}} = u_{\mathrm{L}} = -t_{\mathrm{L}}$ and $v_{\mathrm{R}} = u_{\mathrm{R}} = t_{\mathrm{R}}$, respectively (see figure~\ref{fig_eternal_shock_BH}). In the eternal black hole setting, it is often convenient to use both $u$ and $v$ null coordinates. For completeness, the relevant equations of motion in eqs.~\reef{eq:Y2} and~\reef{dot_v} are reproduced here:
\begin{align} 
\dot{r}_{\pm} &= \pm \sqrt{P_i^2 + r^{2(d-1)} f_i(r) + \frac{4 \tilde{\alpha} P_i^2}{r^2} \left[ f_i^{(0)}(r) - k + P_i^2 r^{-2(d-1)} \right]} \,, \label{eq:rdot_GB_final} \\
\dot{v}_{\pm} &= \frac{1}{f_i(r)} \left( \dot{r}_{\pm} + \frac{P_i}{2 \tilde{\alpha} r^{-2d} \dot{r}_{\pm}^2 - \frac{2 \tilde{\alpha} k}{r^2} - 1} \right) \,, \label{eq:vdot_GB_final} \\
\dot{u}_{\pm} &= \frac{1}{f_i(r)} \left( -\dot{r}_{\pm} + \frac{P_i}{2 \tilde{\alpha} r^{-2d} \dot{r}_{\pm}^2 - \frac{2 \tilde{\alpha} k}{r^2} - 1} \right) \,. \label{eq:udot_GB_final} 
\end{align}
The subscripts $\pm$ denote the branches where the radial coordinate is increasing ($+$) or decreasing ($-$) as we traverse the surface from the left boundary. Note that these expressions also need to be refined to $\mathcal{O}(\tilde{\alpha})$. The same holds for the following analysis.

We trade the boundary conditions $(t_L, t_R)$ for the more tractable set $(P_1, r_s)$, which fully determines the maximal surface. For given values of $P_1$ and $r_s$, we can obtain $\dot{v}_1(r_s)$ and $\dot{r}_1(r_s)$ by evaluating eq.~\reef{dot_v} at $r=r_s$ with $i=1$:
\begin{align}
\dot{r}_1(r_s) = &\, \pm \sqrt{P_1^2 + r_s^{2(d-1)} f_1(r_s) + \frac{4 \tilde{\alpha} P_1^2}{r_s^2} \left[ f_1^{(0)}(r_s) - k + P_1^2 r_s^{-2(d-1)} \right]} \,, \label{eq:rdot1_GB} \\
\dot{v}_1(r_s) = &\, \frac{1}{f_1(r_s)} \left( \dot{r}_1(r_s) + \frac{P_1}{2 \tilde{\alpha} r_s^{-2 d} \dot{r}_1(r_s)^2-\frac{2 \tilde{\alpha} k}{r_s^2}-1} \right) \,. \label{eq:vdot1_GB}
\end{align}
From eq.~\reef{eq:Delta_v_explicit}, the discontinuity in $\dot{v}$ across the shell reads:
\begin{equation}
  \dot{v}_{2} = \dot{v}_{1} +\frac{\tilde{\alpha} \dot{v}^{(0)} \Delta f^{(0)}}{2 r_s^{2(d+1)}} \left[ -2 r_s^{2d} + r_s^2 (\dot{v}^{(0)})^2 \Sigma f^{(0)} \right] \,. \label{eq:Delta_v_explicit_2}
\end{equation} 
As implied by eq.~\eqref{eq:P_i_general}, the conserved momentum jumps across the shock wave. Its value $P_2$ in the $\mathrm{BH}_2$ region evaluates to
\begin{equation}
\begin{aligned}
P_2 &= \frac{r^{2(d-1)} - f_{2} \dot{v}_{2}^2}{2 \dot{v}_{2}} \\
&\quad + \frac{\tilde{\alpha} \Delta_2}{4 r^{2d} (\dot{v}_2^{(0)})^3} \left[ 4k r^{2(d-1)} (\dot{v}_2^{(0)})^2 - \Sigma_2^2 \right] \,, 
\end{aligned}
\label{P_2}
\end{equation}
where
$\Delta_2 \equiv r^{2(d-1)} - f_2^{(0)} (\dot{v}_2^{(0)})^2$ and $\Sigma_2 \equiv r^{2(d-1)} + f_2^{(0)} (\dot{v}_2^{(0)})^2$. 
Depending on the boundary conditions, the maximal surface may traverse either the black or white hole interior of $\mathrm{BH}_1$. These scenarios are dictated by the sign of the initial momentum. Specifically, $P_1=0$ represents the surface crossing bifurcation slice. Potential turning points $r_{t,i}$ in each region are determined by the condition $\dot{r}=0$, yielding
\begin{equation}
P_i^2 + r_{t,i}^{2(d-1)} f_i(r_{t,i}) + \frac{4 \tilde{\alpha} P_i^2}{r_{t,i}^2} \left[ f_i^{(0)}(r_{t,i}) - k + P_i^2 r_{t,i}^{-2(d-1)} \right] = 0 \,.\end{equation} 
We emphasize that a turning point in $\mathrm{BH}_1$ is a prerequisite for any surface entering the white hole sector.
With the parameters $r_s$ and $P_1$ fixed, the calculation of the boundary times $t_\mathrm{L}$ and $t_\mathrm{R}$ reduces to a straightforward integration problem. To streamline the integrations, we define a function $\tau_i[P, r]$ for each region ($i \in \{1,2\}$):
\begin{equation}\label{tau}
\tau_i \left[P_i,r\right] \equiv \frac{1}{f_i(r)} -\frac{P_i}{f_i(r)\sqrt{-\mathcal{U}_i\left(P_i,r\right)} \left(1+2\tilde{\alpha} r^{-2 d} \mathcal{U}_i\left(P_i,r\right)+\frac{2\tilde{\alpha} k}{r^2}\right)} \,,
\end{equation}
where the subscript $i$ distinguishes the spacetime region.
As in Einstein gravity, the time integrals in Gauss-Bonnet gravity maintain the concise forms:
\begin{equation}
\begin{split}
\Delta v_{\pm} &= \int \mathrm{d}v_{\pm} = \int \frac{\dot{v}_{\pm}}{\dot{r}_{\pm}} \mathrm{d}r = \int \tau_i\left[\pm P_i, r\right] \mathrm{d}r \,, \\
\Delta u_{\pm} &= \int \mathrm{d}u_{\pm} = \int \frac{\dot{u}_{\pm}}{\dot{r}_{\pm}} \mathrm{d}r = - \int \tau_i\left[\mp P_i, r\right] \mathrm{d}r \,.
\label{eq:piece_by_piece}
\end{split}
\end{equation} 
The boundary times $t_{\mathrm{R}} + t_w$ and $t_{\mathrm{L}} - t_w$ are subsequently obtained by evaluating eq.~\reef{eq:piece_by_piece} piece-by-piece along the extremal surface. The piecewise integrations are determined by the sign of the momentum $P_1$ and the existence of turning points $r_{t,i}$ in the respective geometries. The explicit piecewise expressions for these boundary times are structurally identical to those derived in~\cite{Chapman:2018lsv}. One simply replaces the Einstein-gravity integrand with our Gauss-Bonnet corrected function $\tau_i[P_i, r]$ defined in eq.~\eqref{tau}. Therefore, we omit rewriting these lengthy piecewise expressions here and directly proceed to the analysis of the complexity growth rate.

The surface may span multiple coordinate patches or cross the shock wave. Although these patches partition the surface and introduce intermediate boundary contributions, the continuity of the generalized momenta $P_\mu$ ensures that they cancel pairwise. At the shock wave, while $\dot{v}$ is discontinuous, the radial equation of motion $\partial_\lambda P_r = \partial_r \mathcal{L}$ guarantees the continuity of the radial momentum across the shell, $[P_r]_{r_s^-}^{r_s^+}=0$. Consequently, the junction contribution $\sim P_r \delta r_s$ vanishes, leaving only the asymptotic boundary terms. The rate of change of the holographic volume thus takes a universal form governed by the boundary momenta:
\begin{equation}
\frac{1}{\Omega_{k,d-1}} \frac{\mathrm{d}W}{\mathrm{d}t} = P_1 \frac{\mathrm{d}t_{\mathrm{L}}}{\mathrm{d}t} + P_2 \frac{\mathrm{d}t_{\mathrm{R}}}{\mathrm{d}t} \,. 
\label{VtshocksF}
\end{equation}
Restricting to the symmetric time evolution ($t_{\mathrm{L}}=t_{\mathrm{R}}=t/2$), this rate simplifies to
\begin{equation}
\frac{\mathrm{d}\cv}{\mathrm{d}t} = \frac{\Omega_{k,d-1}}{2 G_{\mathrm{bulk}} L} \left(P_1+P_2\right) \,. 
\label{CtshocksF}
\end{equation}
While the full evolution requires numerical solutions, the growth rate exhibits distinct analytic behaviors in three characteristic regimes.

\textbf{Late-time limit:}
As $t \to \infty$, the boundary time integrals become divergent. This divergence dictates that the extremal surface asymptotically approaches the critical slices, where the effective potentials and their derivatives vanish simultaneously:
\begin{equation}
\begin{split}
&\partial_r \!\left[\mathcal{U}_1(P_{1,\mathrm{m}},r_{1,\mathrm{m}})\right]=0\,, \qquad
\mathcal{U}_1(P_{1,\mathrm{m}},r_{1,\mathrm{m}})=0,
\\
&\partial_r \!\left[\mathcal{U}_2(P_{2,\mathrm{m}},r_{2,\mathrm{m}})\right]=0\,, \qquad
\mathcal{U}_2(P_{2,\mathrm{m}},r_{2,\mathrm{m}})=0.
\end{split}
\end{equation} 
In this limit, the momenta $P_{1,2}$ approach their maximal critical values $P_{1,2,\mathrm{m}}$. Solving these conditions in perturbative Gauss-Bonnet gravity yields
\begin{equation}\label{maxfaceplanar}
r_{i,\mathrm{m}} = \frac{r_{h,i}}{2^{1/d}}\left(1+\frac{\tilde{\alpha}}{d L^2}\right) \,, \qquad
P_{i,\mathrm{m}} = \frac{r_{h,i}^d}{2 L}\left( 1 - \frac{\tilde{\alpha}}{2 L^2} \right) \,. 
\end{equation}
Substituting these critical momenta into eq.~\eqref{CtshocksF} yields the stable late-time linear growth rate
\begin{equation}
\frac{\mathrm{d}\cv}{\mathrm{d}t} = \frac{4 \pi (M_1+M_2)}{d-1}\left( 1 - \frac{\tilde{\alpha}}{2 L^2} \right) \,. 
\label{lim1}
\end{equation}

\textbf{Intermediate plateau:}
If the shock is injected early enough, a plateau emerges around $t=0$. In this regime, the extremal surface stays near the respective critical slices. Consequently, the growth rate is constant:
\begin{equation}
\frac{\mathrm{d}\cv}{\mathrm{d}t} = \frac{4 \pi (M_2-M_1)}{d-1}\left( 1 - \frac{\tilde{\alpha}}{2 L^2} \right) \,. 
\label{lim2}
\end{equation}

\textbf{Early time regime:}
In the limit evaluated shortly after the shock insertion ($t_{\mathrm{R}}\approx - t_w$), the system enters the early time regime. Here, one still finds $P_1 \approx -P_{1,\mathrm{m}}$, but the second momentum satisfies $P_2 \approx P_{2,\mathrm{m}} - 2 P_{1,\mathrm{m}}$ due to the jump conditions across the shock. Consequently, the growth rate reads:
\begin{equation}
\frac{\mathrm{d}\cv}{\mathrm{d}t} = \frac{4 \pi (M_2 - 3 M_1)}{d-1}\left( 1 - \frac{\tilde{\alpha}}{2 L^2} \right) \,. 
\label{lim3}
\end{equation}
Notably, in the limit $M_1 \to M_2$, these three limits perfectly recover the unperturbed eternal black hole results. Evidently, the stringy effects suppress these growth limits by a factor of $(1 - \tilde{\alpha}/2L^2)$, as seen in eqs.~\reef{lim1}--\reef{lim3}. Beyond this, we shall demonstrate that these corrections further prolong the plateau duration.

\paragraph{Critical time and plateau duration:}
Operationally, we define the end of the plateau regime as the moment the extremal surface in $\mathrm{BH}_1$ traverses the bifurcation surface, transitioning from the white hole to the black hole interior. This event defines a critical time $t_{c,\mathrm{v}}$, characterized by $P_1=0$ and $P_2 \approx P_{2,\mathrm{m}}$. 
Solving the momentum matching condition in eq.~\reef{P_2} perturbatively to $\mathcal{O}(\tilde{\alpha})$ yields the critical value of $r_s$
\begin{equation}
  r_{s,c} = r_{s,c}^{(0)} \left[ 1 + \frac{\tilde{\alpha}}{d L^2} \left( \frac{r_{h,2}^d - r_{h,1}^d}{r_{h,2}^d} \right)^2 \right] \,,
  \label{rsc}
\end{equation}
where $r_{s,c}^{(0)}$ denotes the standard Einstein gravity result:
\begin{equation}
  r_{s,c}^{(0)} = \frac{r_{h,2}^{2}}{(2 r_{h,2}^d - r_{h,1}^d)^{1/d}} \,.
  \label{rsc0}
\end{equation}
Notably, this $\mathcal{O}(\tilde{\alpha})$ correction scales with the square of the mass difference, implying that a Gauss-Bonnet coupling universally pushes $r_{s,c}$ outward toward the boundary. Analyzing $t_{\mathrm{L}}-t_{\mathrm{w}}$ at $P_1=0$ yields
\begin{equation}\label{cvShotc}
t_{\mathrm{L},c} = t_{c,\mathrm{v}}/2 = -v_s + r_1^*(r_{s,c}) \,. 
\end{equation}

The tortoise coordinate $r_1^*(r)$ defined in eq.~\reef{tortoise_coordinates} and the critical shock wave radius $r_{s,c}$ in eq.~\reef{rsc} can be expressed as:
\begin{align}
  r_1^*(r) &= r_1^{*(0)}(r) +\tilde{\alpha}\delta r_1^*(r) \,,
  \label{tortoise_pert}\\
  r_{s,c} &= r_{s,c}^{(0)} + \tilde{\alpha} \delta r_{s,c}=r_{s,c}^{(0)} (1 + \Delta_{\mathrm{GB}}) \,.
  \label{rsc_alpha_corr}
\end{align}
Evaluating the tortoise coordinate at this perturbed critical radius necessitates a consistent expansion to strictly linear order in $\tilde{\alpha}$:
\beq
\begin{split}
r_1^*(r_{s,c}) &= r_1^*(r_{s,c}^{(0)} + \tilde{\alpha} \delta r_{s,c}) \\
&\approx r_1^{*(0)}(r_{s,c}^{(0)} + \tilde{\alpha} \delta r_{s,c}) + \tilde{\alpha} \delta r_1^*(r_{s,c}^{(0)} + \tilde{\alpha} \delta r_{s,c}) \\
&= \left[ r_1^{*(0)}(r_{s,c}^{(0)}) + \frac{\mathrm{d}r_1^{*(0)}}{\mathrm{d}r}\bigg|_{r_{s,c}^{(0)}} \cdot (\tilde{\alpha} \delta r_{s,c}) \right] + \tilde{\alpha} \delta r_1^*(r_{s,c}^{(0)}) \,. 
\end{split}
\label{Double_Expansion}
\eeq
Collecting these pieces, the critical plateau duration $t_{c,\mathrm{v}}$ including the first-order $\tilde{\alpha}$ correction cleanly resolves to
\begin{equation}
t_{c,\mathrm{v}} = 2 t_w + 2 \left[ r_1^{*(0)}(r_{s,c}^{(0)}) + \tilde{\alpha}\left(\frac{1}{f_1^{(0)}(r_{s,c}^{(0)})} \delta r_{s,c} + \delta r_1^*(r_{s,c}^{(0)})\right) \right] \,. 
\label{tc_after_Double_Expansion}
\end{equation}

\textbf{Heavy shock ($w \to \infty$):}
In the heavy shock limit $w \to \infty$ (implying $r_{h,2} \gg r_{h,1}$), the perturbed critical radius in eq.~\reef{rsc_alpha_corr} simplifies to
\beq
r_{s,c} \approx \frac{r_{h,2}}{2^{1/d}} \left( 1 + \frac{\tilde{\alpha}}{d L^2} \right) \,.
\eeq
Physically, a heavy shock strongly backreacts on the background geometry. To satisfy the jump condition across the shell, $r_{s,c}$ is pushed toward the boundary ($r \to \infty$). In this asymptotic regime, the tortoise coordinate expands as
\beq
r_1^*(r) = -\frac{L^2}{r} + \frac{\tilde{\alpha}}{r} = -\frac{L^2}{r} \left( 1 - \frac{\tilde{\alpha}}{L^2} \right) \,.
\eeq
Evaluating this at $r_{s,c}$ yields
\beq
r_1^*(r_{s,c}) \approx -\frac{L^2 \left( 1 - \frac{\tilde{\alpha}}{L^2} \right)}{ \frac{r_{h,2}}{2^{1/d}} \left( 1 + \frac{\tilde{\alpha}}{d L^2} \right) } 
= - \frac{d \cdot 2^{1/d}}{4\pi T_2} \left( 1 - \frac{\tilde{\alpha}}{L^2} \frac{d+1}{d} \right) \,,
\eeq
where we utilized the temperature relation $T_2 = d r_{h,2}/(4\pi L^2)$ from eq.~\reef{temperature_entropy}. Consequently, the critical time eq.~\reef{tc_after_Double_Expansion} evaluates to
\beq
t_{c} = 2t_w - \frac{d \cdot 2^{1/d}}{2\pi T_2} \left( 1 - \frac{d+1}{d} \frac{\tilde{\alpha}}{L^2} \right) \,.
\label{eq:critical_time_heavy_shock}
\eeq

\textbf{Light shock ($w \to 1$):}
In the light shock limit ($r_{h,2} \to r_{h,1}$), the critical radius $r_{s,c}$ approaches the horizon $r_{h,1}$. This behavior aligns with physical intuition: a light shock corresponds to a vanishing shock wave energy $E \to 0$. In the absence of a shock wave ($E=0$), the geometry reduces to that of a pure eternal black hole. 
For the unperturbed eternal black hole ($E = 0$), the extremal surface crosses the bifurcation surface exactly at $r_{h,1}$ ($r_{h,1}=r_{h,2}$), naturally forcing $r_{s,c} \to r_{h,1}$.
Evaluating the perturbative tortoise coordinate in eq.~\reef{tortoise_pert} at this radius, we note that the $\mathcal{O}(\tilde{\alpha})$ correction is regular at the horizon (contributing only a finite constant $\tilde{\alpha}/r_{h,1}$), meaning the logarithmic divergence stems entirely from the zeroth-order Einstein term $r_1^{*(0)}(r_{s,c}) \approx (4\pi T_1)^{-1} \log[(r_{s,c} - r_{h,1})/L_0]$, where $L_0$ is an arbitrary length scale originating from the integration constant of the tortoise coordinate.
We now analyze the displacement of the critical radius from the horizon, $\delta r = r_{s,c} - r_{h,1}$. Using the result in eq.~\reef{rsc_alpha_corr}, the total displacement is
\beq
r_{s,c} - r_{h,1} = \left( r_{s,c}^{(0)} - r_{h,1} \right) + r_{s,c}^{(0)} \Delta_{\mathrm{GB}} \,. 
\eeq
As $w \to 1$, both the Einstein displacement and the Gauss-Bonnet correction factor $\Delta_{\mathrm{GB}}$ scale quadratically with $(w-1)$. Thus, the total displacement scales as a second-order small quantity:
\beq
r_{s,c} - r_{h,1} \approx C \cdot (w-1)^2 \,,
\label{second_order_small_quantity}
\eeq
where $C\equiv d \,r_{h,1} (1 + \tilde{\alpha}/L^2)$ is a constant depending on $\tilde{\alpha}$ but independent of $w$. Substituting this into the logarithmic term yields
\beq
r_1^{*(0)}(r_{s,c}) \approx \frac{1}{4\pi T_1} \log \left[ \frac{C (w-1)^2 }{L_0}\right] = \frac{1}{2\pi T_1} \log(w-1) + \frac{1}{4\pi T_1} \log \frac{C}{L_0}\,.
\eeq
Finally, the critical time in eq.~\reef{tc_after_Double_Expansion} becomes
\beq
t_{c} \approx 2t_w + \frac{1}{\pi T_1} \log(w-1)+ \frac{1}{2\pi T_1} \log \frac{C}{L_0}+2\frac{\tilde{\alpha}}{r_{h,1}} = 2(t_w - t_{\mathrm{scr}}^*) + \text{const.} + \delta t_c \,,
\label{eq:critical_time_light_shock}
\eeq
where the $\text{const.}$ term denotes the finite, $w$-independent term present in pure Einstein gravity, and the scrambling time $t_{\mathrm{scr}}^*$ retains its universal form identical to Einstein gravity:
\beq
t_{\mathrm{scr}}^* = \frac{1}{2\pi T}\log\frac{1}{w-1}= \frac{1}{2\pi T_1} \log S \,. 
\label{scr_S}
\eeq
This result explicitly demonstrates that scrambling is an intrinsic property of the near-horizon geometry. It is governed exclusively by the horizon temperature $T_1$ and remains completely insensitive to higher-derivative Gauss-Bonnet modifications in the planar case.
By isolating the deviation from the Einstein gravity result, we obtain $\delta t_c = t_c^{(\mathrm{GB})} - t_c^{(\mathrm{Ein})} = \tilde{\alpha} [ (2\pi T_1 L^2)^{-1} + 2/r_{h,1} ]$.
The absence of $w$ in this expression demonstrates that the prolongation of the critical duration caused by stringy corrections originates entirely from the pure background effect.

We now verify whether the entropy $S_{\mathrm{match}}$ derived from $t^*_{\mathrm{scr}}$ is consistent with the entropy $S\sim r_h^{\,d-1}$ given in eq.~\reef{temperature_entropy} for $k=0$. To relate this to thermodynamic quantities, recall that $w = r_{h,2}/r_{h,1}$. Since $M \propto r_h^d$, the mass ratio expands as $M_2/M_1 = w^d \approx 1 + d(w-1)$. The injected shock wave energy is thus $E = M_2 - M_1 \approx d M_1 (w-1)$, yielding the perturbation parameter $w-1 \approx E / (d M_1)$. From eq.~\reef{scr_S}, the effective entropy is identified as $S_{\mathrm{match}} \sim (w-1)^{-1}$. Imposing the minimal perturbation scale $E \sim T$, we perfectly recover the thermodynamic scaling: $S_{\mathrm{match}} \sim M/T \sim r_h^d / r_h \sim S$.
This matching is thermodynamically consistent. For the planar case, the Gauss-Bonnet term modifies neither the temperature nor the entropy in eq.~\reef{temperature_entropy}. Therefore, the scrambling time derived from the light shock limit perfectly reproduces the thermodynamic relation $t_{\mathrm{scr}}^* \sim T^{-1} \log S$. This confirms that holographic complexity serves as a sensitive probe of microscopic degrees of freedom, faithfully capturing the scrambling behavior even in the presence of higher-curvature corrections.

\begin{figure}[htbp]
\centering
\includegraphics[width=0.48\textwidth]{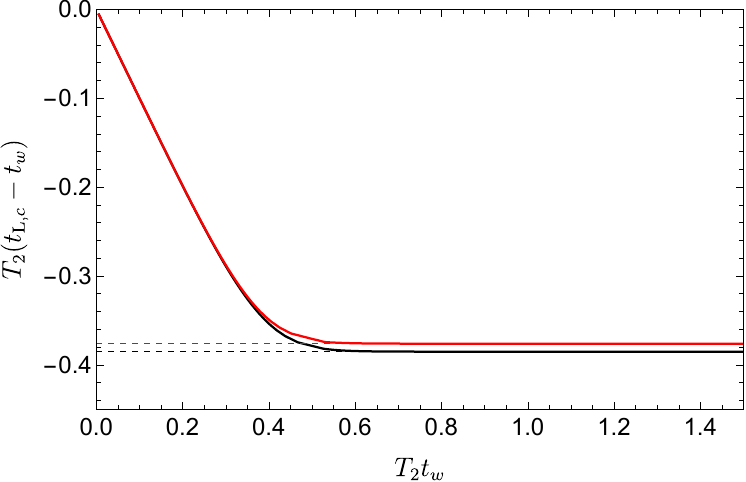}
\hfill \includegraphics[width=0.48\textwidth]{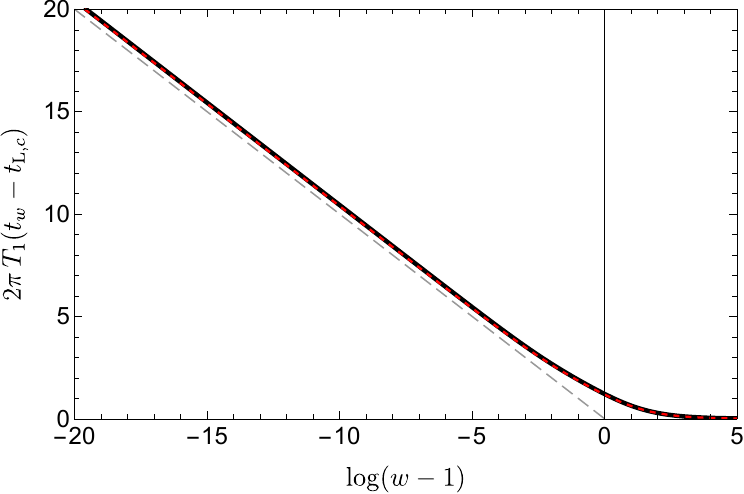}
\caption{The critical time $t_{\mathrm{L},c}$ for $d=4$ planar black holes. Left: The heavy shock regime ($w=2$). The scaled critical time $T_2(t_{\mathrm{L},c}-t_w)$ is plotted against the shock time $T_2 t_w$ for Einstein gravity (black solid) and GB gravity with $\alpha/L^2=1/100$ (red solid). The horizontal dashed lines (gray for Einstein, red for GB) represent the early shock analytical predictions from eq.~\eqref{tc_after_Double_Expansion}.
Right: The light shock regime for an early shock $T_2 t_w = 6$. The logarithmic dependence on $w-1$ illustrates the scrambling time behavior. The dashed gray line depicts a reference slope of $-1$. To ensure visibility, the Gauss-Bonnet result (thin dashed red) is superimposed on the Einstein baseline (thick solid black), demonstrating near-perfect overlap.}
\label{fig:CriticalTime}
\end{figure}

\begin{figure}[htbp]
\centering \includegraphics[width=0.6\textwidth]{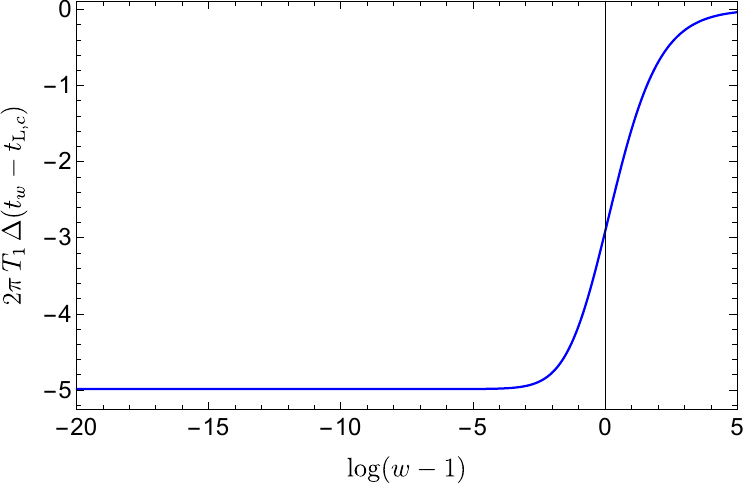}
\caption{The residual difference in the scaled critical time $2\pi T_1(t_w-t_{\mathrm{L},c})$ between Gauss-Bonnet ($\alpha/L^2=1/100$) and Einstein gravity for the light shock regime, extracted from the right panel of figure~\ref{fig:CriticalTime}. Data are scaled by a factor of 100 for clarity. As $w \to 1$, the difference saturates to a distinct constant, confirming that the Gauss-Bonnet correction strictly shifts the scrambling time without altering its logarithmic divergence.}
\label{fig:ScramblingDiff}
\end{figure}

We evaluate the full numerical evolution of the critical time for $d=4$ Gauss-Bonnet gravity ($\alpha/L^2=1/100$), focusing on two characteristic regimes. For a heavy shock ($w=2$, left panel of figure~\ref{fig:CriticalTime}), the numerical curves at early injection times (large $T_2 t_w$) precisely asymptote to the constant limits predicted by eq.~\eqref{tc_after_Double_Expansion}. Physically, a Gauss-Bonnet coupling strictly prolongs this plateau duration relative to the Einstein gravity baseline. In the light shock limit ($w \to 1$, right panel of figure~\ref{fig:CriticalTime}), the critical time serves as a precise indicator for scrambling. Both theories exhibit the expected scrambling behavior.
Since the $\mathcal{O}(\alpha)$ shift is visually indistinguishable on the logarithmic scale of figure~\ref{fig:CriticalTime}, we isolate and magnify this residual difference in figure~\ref{fig:ScramblingDiff}. As $w \to 1$, the difference cleanly asymptotes to a constant value. This provides direct numerical verification of our analytical deduction: higher-curvature corrections shift the scrambling time, leaving the logarithmic coefficient strictly unmodified.

\section{Summary}\label{sec:Discussion}
In this work, we have investigated the full time dependence and late-time behavior of holographic complexity in Gauss-Bonnet gravity utilizing the complete volume proposal. By evaluating the complexity growth in both static and dynamic setups, we uncovered several stringy corrections to the pure Einstein gravity baseline.

For the unperturbed eternal black hole, we established that the complexity growth rate is modulated by both the curvature and the Gauss-Bonnet coupling, as shown in eq.~\reef{eq:expansion_of_late_time_limit2}. Ultimately, the interplay between higher-order $\hat{\alpha}$ terms and curvature corrections dictates the upper and lower bounds of the complexity growth rate for spherical and hyperbolic black holes relative to their counterparts in the $k = 0$ case. Notably, stringy corrections preserve the suppression in the planar case and drive the final slice closer to the horizon, as shown in eqs.~\reef{eq:expansion_of_late_time_limit2} and~\reef{eq:rf_expansion}. For non-planar topologies, the series expansion explicitly reveals a nontrivial interplay between curvature and stringy corrections that dictates whether the late-time growth rate experiences an anomalous acceleration or a reinforced suppression. Furthermore, a universal ``competition effect'' emerges for all $k$, as the Gauss-Bonnet correction transitions from accelerating early-time growth to suppressing late-time evolution. This reveals a nontrivial feature distinct from the standard CV proposal.

For the one-sided Vaidya spacetime formed by a collapsing null shell, our perturbative analysis in the small Gauss-Bonnet coupling limit reveals a notable modification: the junction conditions across the null shell necessitate jumps in the derivatives of both the Eddington-Finkelstein time ($\dot{v}$) and the radial coordinate ($\dot{r}$). We demonstrated that the rate of complexity growth is directly proportional to the conserved momentum $P_{\mathrm{BH}}$, as shown in eq.~\reef{VolVadFinal}. Specifically, for planar horizons ($k=0$), while this conserved momentum remains constant as in the Einstein gravity case, its magnitude is explicitly suppressed by the Gauss-Bonnet corrections. For the spherical case ($k=1$), upon incorporating stringy effects, the rate of growth still decreases at early times and approaches the late-time limit from above. However, while these stringy corrections are negligible at early times, they significantly alter the late-time limit.

Finally, our investigation into two-sided Vaidya spacetimes perturbed by shock waves provided a focused analysis on the critical time and the plateau duration. From eqs.~\reef{eq:critical_time_heavy_shock} and~\reef{eq:critical_time_light_shock}, we observe that for both heavy and light shocks, the leading $2t_w$ scaling naturally reflects that an earlier shock injection leads to a prolonged duration of the switchback effect. Moreover, the stringy correction explicitly prolongs the critical time $t_c$. As the stringy effects become stronger, the duration of the switchback cancellation is further extended, yet the underlying scrambling behavior of the system remains preserved. Additionally, since the planar horizon temperature $T_1 = d r_{h,1}/(4\pi L^2)$ receives no $\mathcal{O}(\tilde{\alpha})$ corrections, the Lyapunov exponent $\lambda_L = 2\pi T_1$ remains invariant, which is consistent with the discussion in~\cite{Shenker:2013pqa}.

\subsection*{Future Directions}
Several intriguing avenues for future research remain. First, while our perturbative analysis of the one-sided Vaidya spacetime elucidated the underlying dynamics, extending this framework to hyperbolic black holes ($k=-1$) could reveal novel topological effects on the complexity growth rate during the collapse. Second, regarding two-sided black holes perturbed by shock waves, our current study focused exclusively on planar horizons ($k=0$). Generalizing this to non-planar topologies and explicitly constructing the full time-evolution profiles of complexity for such configurations would be highly valuable. Third, our treatment of dynamical setups relied heavily on a small Gauss-Bonnet coupling expansion. Moving beyond this perturbative regime to exact, non-perturbative analyses in dynamical higher-derivative backgrounds presents a challenging but crucial next step. Additionally, evaluating the complete volume proposal within other higher-curvature theories warrants further investigation. Finally, our exact bulk analysis of the switchback effect provides a rigorous geometric target for boundary proposals. Exploring how these dynamical shock wave scenarios emerge from the spectral generating functions of perturbed TFD states, as recently envisioned by~\cite{Miyaji:2025jxy}, remains a compelling open challenge. Investigating these subtleties will undoubtedly deepen our understanding of the holographic dictionary for complexity in higher-derivative gravity.

\section*{Acknowledgments}
We would like to thank Shan-Ming Ruan and Hong-Yue Jiang for their very useful discussions and valuable comments on an earlier draft of this paper. We also thank Chao-Xi Fang, Jing-Cheng Chang and Shan-Ping Wu for their selfless help. 
This work was supported by the National Natural Science Foundation of China (Grants No. 12475056 and No. 12247101), 
the Fundamental Research Funds for the Central Universities (Grant No. lzujbky-2025-jdzx07), 
the Natural Science Foundation of Gansu Province (No. 22JR5RA389, No. 25JRRA799), 
the 111 Project (Grant No. B20063) and Gansu Province's Top Leading Talent Support Plan. 
\appendix
\section{Analytical Branches of the Effective Potential}
\label{appendix:A}
Recall that the effective potential is identically defined as $\mathcal{U} = -\dot{r}^2$. To present its exact analytical branches in a compact form, we first solve the cubic equation~\eqref{eq:cubic_EOM} for $\dot{r}^2$ by defining the following auxiliary functions:
\begin{align}
  U(r) &= r^2 + 2k\tilde{\alpha} + \tilde{\alpha}f(r) \,, \\
  V(r) &= r^2 + 2k\tilde{\alpha} - 2\tilde{\alpha}f(r) \,, \\
  H(r) &= 27 P^2 \tilde{\alpha} r^{6-2d} - V(r)^3 \,, \\
  Z(r) &= H(r) + \sqrt{H(r)^2 - V(r)^6} \,.
\end{align}
By introducing $\omega = (-1 + i\sqrt{3})/2$, the three roots $\dot{r}_i^2$ ($i=1,2,3$) obtained via Cardano's formula can be elegantly expressed as:
\begin{align}
  \dot{r}_1^2 &= \frac{r^{2d-2}}{6 \tilde{\alpha}} \left[ 2 U(r) + \frac{V(r)^2}{Z(r)^{1/3}} + Z(r)^{1/3} \right] \,, \label{eq:Y1} \\[6pt]
  \dot{r}_2^2 &= \frac{r^{2d-2}}{6 \tilde{\alpha}} \left[ 2 U(r) + \omega^2 \frac{V(r)^2}{Z(r)^{1/3}} + \omega Z(r)^{1/3} \right] \,, \label{eq:Y2} \\[6pt]
  \dot{r}_3^2 &= \frac{r^{2d-2}}{6 \tilde{\alpha}} \left[ 2 U(r) + \omega \frac{V(r)^2}{Z(r)^{1/3}} + \omega^2 Z(r)^{1/3} \right] \,. \label{eq:Y3}
\end{align}
\section{Expansion Coefficients for the Growth Rate}
\label{app:coefficients}

In this appendix, we present the explicit expressions for the higher-order coefficients appearing in the geometric and thermodynamic expansions of the complexity growth rate, eqs.~\eqref{eq:expansion_geometric} and~\eqref{eq:expansion_of_late_time_limit2}.
To simplify the notation, we define the dimension-dependent constants
\begin{equation}
  \kappa \equiv 4^{1/d} \,, \qquad \xi \equiv 2^{2/d} \,, \qquad \text{and} \quad \tilde{a} \equiv \frac{\kappa \bigl( (\xi^{-1} - 1) d + 1 \bigr) k}{d^2} \,.
\end{equation}
The mass correction factor $\mathcal{B}_d$, relevant for the hyperbolic geometry ($k=-1$), is given by
\begin{align}
  \mathcal{B}_d &= -2^{1-\frac{d}{2}} d^{d/2}
  \left( \sqrt{(d-2)^2 - 4 \hat{\alpha} (d-4) d} + d - 2 \right)^{\frac{d-4}{2}} \nonumber \\
  &\quad \times \left[ 2 \hat{\alpha} d \bigl( d (k^2 - 1) + 4 \bigr) + (d k + d - 2) \left( \sqrt{(d-2)^2 - 4 \hat{\alpha} (d-4) d} + d - 2 \right) \right] \,.
\end{align}
The subleading coefficients read
\begin{align}
  \mathcal{A}_2 &= -2^{\frac{2}{d}-1} k \left( 1 - \frac{3d+4}{2d} \hat{\alpha} \right) + \mathcal{O}(\hat{\alpha}^2) \,, \\
  \mathcal{C}_2 &= -2^{\frac{2}{d}-1} d^2 k \left( 1 - \frac{3d+4}{2d} \hat{\alpha} \right) + \mathcal{O}(\hat{\alpha}^2) \,. \label{eq:C2}
\end{align} 
The fourth-order coefficients $\mathcal{A}_4$ (geometric expansion) and $\mathcal{C}_4$ (thermodynamic expansion) are decomposed into the Einstein gravity contribution (zeroth order in $\hat{\alpha}$) and the Gauss-Bonnet correction (first order in $\hat{\alpha}$):
\begin{equation}
  \mathcal{A}_4 = \mathcal{A}_4^{(0)} + \hat{\alpha} \mathcal{A}_4^{(1)} + \mathcal{O}(\hat{\alpha}^2) \,, \qquad
  \mathcal{C}_4 = \mathcal{C}_4^{(0)} + \hat{\alpha} \mathcal{C}_4^{(1)} + \mathcal{O}(\hat{\alpha}^2) \,.
\end{equation}
The coefficients for the geometric expansion ($\mathcal{A}_4$) are
\begin{align}
\mathcal{A}_4^{(0)} &= -\frac{1}{4} \tilde{a}^2 d^2 - \frac{1}{16} \left( 4 - 8\xi + \kappa^2 \right) k^2 + \frac{1}{2}\tilde{a} \left( -\kappa d + d + \kappa \right) k \,, \\
  \mathcal{A}_4^{(1)} &= -\frac{1}{16d} \Bigl[ 28 \tilde{a}^2 d^3 + \bigl( (28 - 40\xi + 27\kappa^2) d + 8\xi (4 - \kappa) \bigr) k^2 \nonumber \\
  &\quad\quad + 8 \tilde{a} \bigl( (-7 + 5\kappa) d^2 - \kappa d + 4\kappa \bigr) k \Bigr] \,.
\end{align}
The coefficients for the thermodynamic expansion ($\mathcal{C}_4$) are
\begin{align}
  \mathcal{C}_4^{(0)} &= \frac{d^3}{16} \Bigl[ -\kappa^2 d k^2 - 8\xi \left( \tilde{a} (d-1) d - 2k \right) k - 4d \left( k - \tilde{a} d \right)^2 \Bigr] \,, \\
  \mathcal{C}_4^{(1)} &= -\frac{7}{4} d^4 \left( k - \tilde{a} d \right)^2 + \frac{\xi^2 d^3}{16} (8 - 27 d) k^2 \nonumber \\
  &\quad\quad - \frac{1}{2}\kappa d^2 k \Bigl[ \tilde{a} d \bigl( d(5d-1) + 4 \bigr) + 2 \left( -6d^2 + 3d + 4 \right) k \Bigr] \,.
\end{align}

\end{document}